\documentclass[aps,superscriptaddress,prd,
onecolumn,
floatfix, 
nofootinbib,
amsmath,amssymb,amsfonts,longbibliography]{revtex4-2}

\usepackage{mathrsfs}
\usepackage[pdftex]{graphicx}
\usepackage{color,float}
\usepackage[colorlinks,linkcolor=blue,anchorcolor=violet,citecolor=red]{hyperref}
\usepackage{physics}
\usepackage{comment}
\usepackage[normalem]{ulem}
\newcommand{\souteq}[1]{\hbox{}}

\newcommand{\del}{\delta}

\newcommand{\bs}[1]{\boldsymbol{#1}}

\newcommand{\ave}[1]{\langle{#1}\rangle}
\newcommand{\kt}[1]{|{#1}\rangle}
\newcommand{\br}[1]{\langle{#1}|}

\begin{document}

\title{Quantumness of gravity in harmonically trapped particles}

\author{Youka Kaku}
\email{kaku.yuka.g4@s.mail.nagoya-u.ac.jp}
\author{Shin'ya Maeda}
\email{maeda.shinya.k2@s.mail.nagoya-u.ac.jp}
\author{Yasusada Nambu}
\email{nambu@gravity.phys.nagoya-u.ac.jp}
\author{Yuki Osawa}
\email{osawa.yuki.e8@s.mail.nagoya-u.ac.jp}
\affiliation{Department of Physics, Graduate School of Science, Nagoya
University, Chikusa, Nagoya 464-8602, Japan}

\date{December 10, 2022} 
\begin{abstract}
  This study investigates the quantumness of gravity under the setup
  of the atomic interferometry from the viewpoint of mass-energy equivalence. We evaluated interference visibility
  considering a particle with internal energy levels in a harmonic
  trapping potential. As per the result, for a spatially superposed
  gravitational source mass, interference visibility exhibits collapse
  and revival behavior, which implies that an initial separable
  internal state evolves to the entangled state with respect to the
  degrees of freedom of the center of mass, the internal energy
  levels, and the external source state. In particular, it does not
  exhibit revival behavior when gravity is treated as a quantum
  interaction, while it revives with a finite period for a 
  semiclassical treatment of gravity. We also examined the temporal
  behavior of entanglement negativity and found that the nonrevival behavior of visibility reflects the creation of the entanglement
  between the internal energy state and the external source state which is
  uniquely induced by the quantum interaction of gravity in accordance
  with the weak equivalence principle.
\end{abstract}

\maketitle

\section{Introduction}

The unification of gravity and quantum theory is one of the most
challenging subjects in modern physics because the experiment to test
quantum gravity has not been realized thus far. Nevertheless, there
are several experiments that focus on the quantum aspect of a particle
under the classical external gravitational field to examine the
relation between quantum mechanics and classical gravity
\cite{Sakurai2011}. For example, in the Colella-Overhauser-Werner
(COW) experiment \cite{Colella1975}, the difference in gravitational
potential causes the phase difference of a quantum particle and leads
to gravity-induced quantum interference. This phenomenon was confirmed
in the experiment using a neutron interferometer \cite{Rauch2000}.

As a further development of quantum experiments under classical
gravity, the authors of paper~\cite{Pikovski2015} discuss the
decoherence mechanism of a quantum particle with internal (INT)
degrees of freedom using mass-energy equivalence. Herein, the particle
with the INT degrees of freedom is characterized by two systems:
center of mass (CM) system and INT system. According to mass-energy
equivalence, the particle acquires a different mass depending on the
INT energy levels. Therefore, the dynamical variables of the CM and
the INT systems entangle via special relativistic and gravitational
couplings, which is known as universal gravitational decoherence. As a
result, when atomic interferometry is considered, the interference
visibility of the INT state exhibits collapse and revival
behavior. Although time dilation due to classical gravity in a
  quantum clock system is
  recently measured  in a feasible laboratory
  experiment \cite{Bothwell2022}, universal gravitational
  decoherence has not been confirmed yet. In addition, details
regarding the entanglement behavior have been studied in
~\cite{Zych2016,Zych2019}, and the gravitational decoherence in the
Ramsey interferometry is also discussed by Haustein \textit{et
  al.}~\cite{Haustein2019}.

Although the COW experiment \cite{Colella1975} and the gravitational decoherence introduced
by Pikovski \textit{et al.} \cite{Pikovski2015} treat the quantum system under classical
external gravity, they focus  on the quantum aspects of the probe
particle rather than the quantumness of gravity.  Recently, as the first step to
tackle the quantumness of gravity, ideas based on quantum information
have been proposed to test the quantumness of low-energy Newtonian
gravity in tabletop experiments. These ideas are referred to as
 the Bose-Marletto-Vedral (BMV) proposal
\cite{Bose2017,Marletto2017,Christodoulou2019}, and are  based on the
principle of quantum information theory, which states that local operations and
classical communication (LOCC) cannot create  entanglement between
two systems \cite{Nielsen2007}. Based on this principle,  gravity can
be clarified as a quantum interaction or not by detecting
the creation of gravity-induced entanglement. The BMV proposal
received a lot of attention, and has stimulated many other related
proposals \cite{Carney2019,Matsumura2020,Matsumura2022}.  In  interferometry
experiments, the creation of entanglement between the probe and the
environment is reflected in the quantum decoherence of the probe
system \cite{Feynman2000,Joos1985,Zurek2003}. Carney \textit{et al.}
\cite{Carney2021} explored the quantum gravity-induced decoherence in
a hybrid system that comprised a massive oscillator and a source mass
particle in cat state, while some comments on this paper are discussed in \cite{Streltsov2022} that LOCC with stochastic noise in the system can also reproduce the decoherence.

We note here that the purpose of the BMV proposal and other related proposals is to confirm if gravity produces quantum entanglement in the nonrelativistic scale. These proposals just assume that the Newtonian gravitational potential is treated as a two-body operator between gravitational sources, which is the consequence of the quantum field theory to explain the gravitational force by exchanging gravitons.
Even though gravity-induced entanglement does not lead to the quantization of the gravitational field immediately, it is worth a try to explore as the first step to investigating quantum gravity for the following two reasons; 
First, it may be possible to detect gravity-induced entanglement in the near future as another way to approach the quantum feature of gravity based on the recent progress in the quantum experiments of macroscopic objects. For instance, since we need to treat the Planck energy scale under strong gravity, such as nearby the blackhole, to detect graviton, it is challenging to obtain direct evidence of graviton with terrestrial colliders. On the other hand, gravity-induced entanglement is expected to be the new direction to confirm the quantumness of gravity by focusing on the gravity of the Planck mass object in the tabletop experiment, although it cannot support the existence of graviton directly.
Second, testing the gravity-induced entanglement in the nonrelativistic scale gives a valuable clue to the Newtonian limit of quantum gravity. 
Respecting the above, we will investigate gravity-induced quantum entanglement, which we refer to as the quantumness of gravity throughout this paper.

In this study, we propose a new approach to capture the quantumness of
gravity in the Ramsey interferometry from the point of view of
mass-energy equivalence. We assumed that a probe particle with two
energy levels as the INT degrees of freedom is trapped in the harmonic
oscillator potential, and 
feels external gravity, which mass source is in a cat state.
Based on this setup, we aim to detect the interference visibility of the probe particle in Ramsey interferometry and calculate it for two cases;
when the external gravity produces the entanglement between the probe particle system and the gravitational mass source system, or not. To perform a specific calculation, we adopt a particular form of gravitational interaction. For the former case, we assume the first quantization of Newtonian gravity in Eq.~\eqref{eq:QGpotential}, which contains operators of both the particle and the source system and produces quantum entanglement between them. For the latter case, we assume the semiclassical gravity\cite{Kibble1978,Kibble1980,Diosi1984} which will be introduced in Eq.~\eqref{eq:CGpotential}. Herein, semiclassical means that the gravitational mass source is quantized, but gravity remains fundamentally classical. 
As a result, we will show that the quantumness of gravity is reflected as a
nonrevival behavior of the interference visibility, and that it is
related to the creation of genuine tripartite system entanglement. 
We will also see that under the leading order approximation with respect to the separation of the source,  this nonrevival behavior does not appear for
other quantum interactions such as the electromagnetic Coulomb interaction respecting the weak equivalence principle; Our proposal successfully captures quantum nature unique to gravity.

The structure of the remainder of this paper is as follows. In
Sec.~\ref{sec:Ramsey_interferometry}, we will briefly review the
Ramsey interference~\cite{Ramsey1950}. In Sec.~\ref{sec:internal}, we
explore a particle with INT degrees of freedom in an external
gravitational field.  In Sec. \ref{sec:Main} we introduce the
experimental setup, calculate the transition probability in Ramsey
interference following Haustein \textit{el al.}~\cite{Haustein2019},
and reveal the behavior of the interference visibility under quantum
superposition of gravitational source. In Sec.~\ref{sec:Negativity},
we estimate the creation of entanglement by calculating the
entanglement negativity. Section~\ref{sec:Discussion}, we discuss the
results of the study and explore the comparison of quantized gravity
and other quantum interactions. Section~\ref{sec:Summary} presents a
summary of this study. The unit of $\hbar=1$ was adopted throughout
the study.
\section{Ramsey interferometry}
\label{sec:Ramsey_interferometry}
In this section, we briefly review the concept of Ramsey
interferometry \cite{Ramsey1950,Cronin2009,Abend2019}.  Considering a particle (atom)
with two internal energy levels, the Hamiltonian of the particle can
be expressed as
\begin{equation}
  \hat H=\sum_{j=0,1}\hat H_j\ket{E_j}\bra{E_j},\quad \bra{E_j}\ket{E_k}=\del_{jk},
\end{equation}
where $\ket{E_j}$ denotes the $j$th internal energy eigenstate, and
$\hat H_j$ denotes the CM Hamiltonian of the particle associated with the
$j$th internal energy level. The evolution operator of the total
state is expressed as
\begin{equation}
  {\hat U}(t)=\exp\left[-i\hat H
    t\right]=\sum_j \hat U_j(t)\ket{E_j}\bra{E_j},
  \label{eq:evo}
\end{equation}
where $\hat U_j=\exp\left(-i\hat H_jt\right)$ is the
evolution operator of the particle state with the $j$th internal
energy level.  The Ramsey interferometry is performed as per the
following steps
(Fig. \ref{fig:interf_CG}):
\begin{figure}[H]
  \centering
  \includegraphics[width=0.4\linewidth,clip]{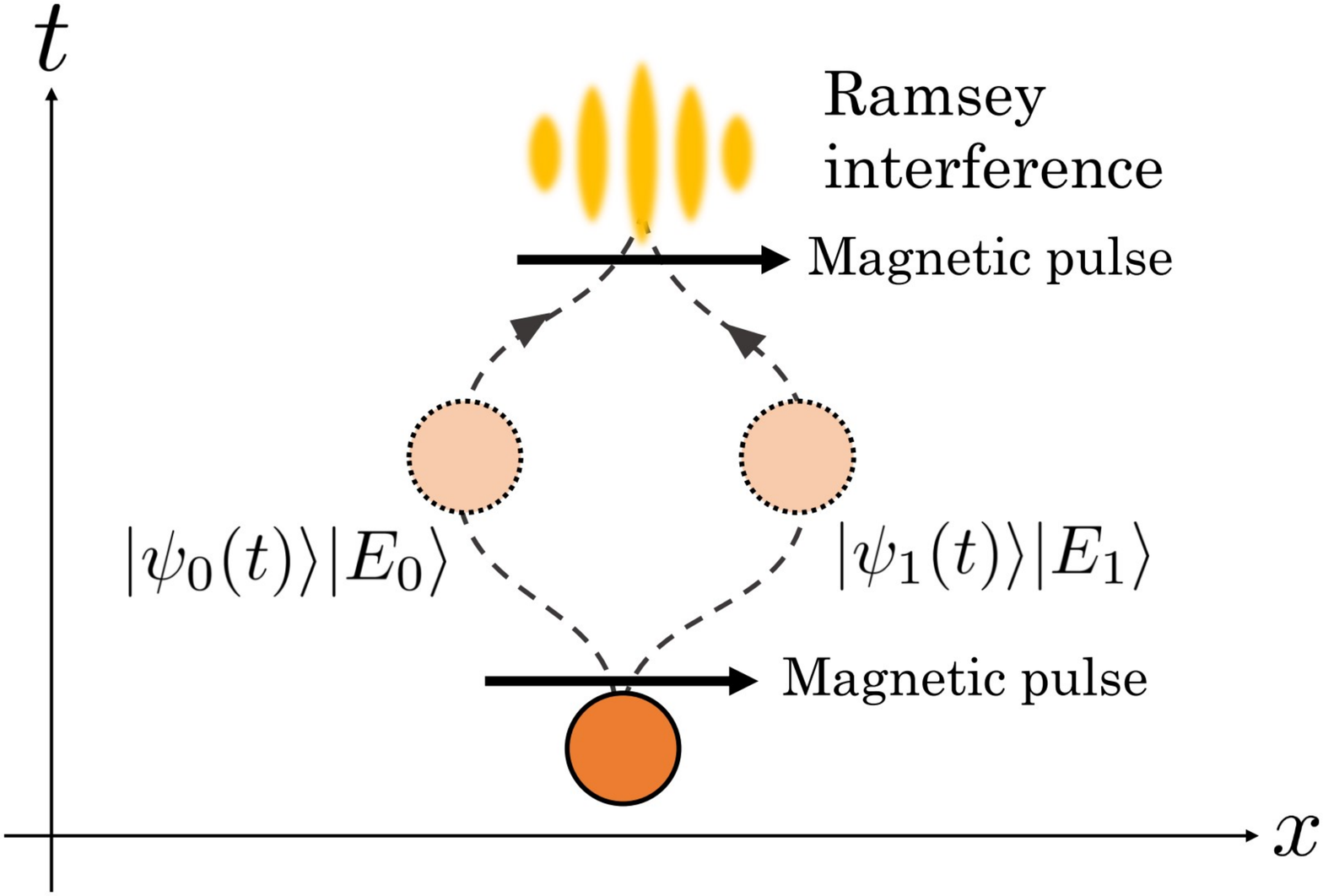}
  \caption{Schematic representation of  Ramsey
    interferometry using a superposed particle with different internal energy
    levels. After evolution, information on the INT state is
    obtained from the interference visibility.}
  \label{fig:interf_CG}
\end{figure}
\begin{description}
  \item [Step 0] Prepare the initial state of the particle  with
  the internal energy $E_0$:
  \begin{equation}
    \ket{\Psi(0)}=\ket{\psi(0)}\otimes\ket{E_0},
  \end{equation}
  where $\ket{\psi(0)}$ denotes the initial CM state.
  \item [Step 1] Apply a $\pi/2$-pulse to take the initial state into
  a superposition of two internal energy
  eigenstates:\footnote{Identical to applying the Hadamard gate
    $H=\frac{1}{\sqrt{2}}\begin{bmatrix}1&1\\1&-1\end{bmatrix}$ to a
    qubit state.}
  \begin{equation}
    \ket{\Psi'(0)}=\ket{\psi(0)}\otimes\left(\ket{E_0}+\ket{E_1}\right)/\sqrt{2}.
  \end{equation}
  \item [Step 2] Evolve the superposed state $\ket{\Psi'(0)}$ with the evolution
  operator mentioned in Eq. \eqref{eq:evo} up to time $t$:
  \begin{align}
    \ket{\Psi(t)}&=\hat U(t)\ket{\Psi'(0)}=\biggl(\hat
                   U_0(t)\ket{\psi(0)}\otimes\ket{E_0}+
                   \hat U_1(t)\ket{\psi(0)}\otimes\ket{E_1}\biggr)/\sqrt{2}.
  \end{align}
  \item [Step 3] Apply the $\pi/2$-pulse again to make the INT state as
  $\ket{E_0}\rightarrow (\ket{E_0}+\ket{E_1})/\sqrt{2},~
  \ket{E_1}\rightarrow (\ket{E_0}-\ket{E_1})/\sqrt{2}$:
  \begin{align}
    \ket{\Psi'(t)}&=\frac{1}{2}\biggl[\hat
                    U_0(t)\ket{\psi(0)}\otimes(\ket{E_0}+\ket{E_1})
                    +\hat
                    U_1(t)\ket{\psi(0)}\otimes(\ket{E_0}-\ket{E_1})\biggr]
                    \notag\\
    &=\frac{1}{2}\biggl[\left(\hat U_0(t)+\hat
      U_1(t)\right)\ket{\psi(0)}\otimes\ket{E_0}
      +\left(\hat U_0(t)-\hat
      U_1(t)\right)\ket{\psi(0)}\otimes\ket{E_1}
      \biggr].
  \end{align}
  \item [Step 4] Measure the occupation probability of the lower
  energy eigenstate $\ket{E_0}$:
  \begin{align}
    P(t)&=\frac{1}{4}\bra{\psi(0)}(\hat U_0^\dag+\hat U_1^\dag)(\hat
           U_0+\hat U_1)\ket{\psi(0)} \notag \\
     &=\frac{1}{2}\left(1+\mathrm{Re}\left[\bra{\psi(0)}\hat
      U_0^\dag(t)\hat U_1(t)\ket{\psi(0)}\right]\right) \notag \\
         &=\frac{1}{2}\bigl(1+\mathrm{Re}\left[\bra{
           \psi_0(t)}\ket{\psi_1(t)}\bigr]\right).
           \label{eq:prob} 
  \end{align}
\end{description}
\noindent
Herein, $|\psi_i(t)\rangle\equiv \hat{U}_i|\psi(0)\rangle$.
Expressing
$\bra{\psi_0(t)}\ket{\psi_1(t)}=|\mathcal{V}(t)|\,e^{i\Theta(t)}$
with real functions $|\mathcal{V}|$ and $\Theta$, the occupation
probability is expressed as
\begin{equation}
  P(t)=\frac{1}{2}\left(1+|\mathcal{V}(t)|\cos\Theta(t)\right).
\end{equation}
Herein, $|\mathcal{V}(t)|$ is the interference visibility, which
contains information regarding the internal energy levels of the
particle.  For example, if $H_i=E_i$, then
$\hat U_j(t)=e^{-iE_jt}$ and
$P(t)=\frac{1}{2}(1+\cos(\Delta E\,t)), \Delta E=E_1-E_0$. 
By measuring this probability as the function of time, we can determine
an energy gap of this system. In this case, the interference
visibility is equal to one.

\section{particle with internal states in the external gravitational
  field}
\label{sec:internal}
We consider a particle with INT degrees of freedom in the
external gravitational field. 
When the particle is moving slowly in a weak gravitational field, any internal energy contributes to the total rest mass of the particle respecting the mass-energy equivalence as follows:
\begin{align}
    \hat m=m_0 \hat I + \hat H_{\text{INT}}/c^2
\end{align}
Here, $m_0$ is the rest mass for the CM system, $H_{\text{INT}}$ is the Hamiltonian for the INT system, and they satisfy $H_{\text{INT}}\ll m_0c^2$. Let us assume the particle to have two
internal energy levels, namely $E_0=0$ and $E_1=E$. Then by substituting
\begin{align}
    \hat H_\text{INT}=\sum_{j=0,1}E_j\ket{E_j}\bra{E_j}, \label{eq:HINT}
\end{align}
the total rest mass of the particle is given as
\begin{equation}
\label{eq:mass}
  \hat m=m_0\,\hat I+\frac{1}{c^2}\sum_{j=0,1}E_j\ket{E_j}\bra{E_j},\quad \hat I=\sum_{j=0,1}\ket{E_j}\bra{E_j}.
\end{equation}
The total state of the
particle system is described by the CM system and the INT system. The
CM system is characterized by its position $\bs{x}$ and conjugate
momentum $\bs{p}$. The INT system is characterized by its
energy eigenstate $\ket{E_j}$. Considering a weak external
gravitational field,  the metric is given by
\begin{equation}
  ds^2=-(1+2\Phi(\bs{x})/c^2)dt^2+(1-2\Phi(\bs{x})/c^2)d\bs{x}^2,
\end{equation}
where $\Phi(\bs{x})$ is the gravitational potential with $|\Phi|/c^2\ll
1$. 
In general, the Hamiltonian of a free-falling particle with its mass $m$ on the background spacetime is
\begin{align}
    H=\sqrt{-g_{00}(m^2 c^4+c^2g_{ij}p^ip^j)}\notag .
\end{align}
This time, since we consider the particle whose mass depends on the internal energy level as in \eqref{eq:mass}, and assume that the system is on a less relativistic scale, we obtain
\begin{align}
    H\approx H_\text{CM}+\left(1+\Gamma(\bs{x},\bs{p})/c^2\right)H_\text{INT},\quad
     \Gamma(\bs{x},\bs{p}):=-\frac{\bs{p}^2}{2m_0^2}+\Phi(\bs{x}) \label{eq:H1}
\end{align}
as a lowest order approximation in Taylor expansion of $1/c^2\ll1$ and $H_{\text{INT}}/(m_0c^2) \ll 1$. Here, $H_\text{CM}$ is the CM Hamiltonian whose explicit form is
\begin{align}
  H_\text{CM}&=m_0+\frac{\bs{p}^2}{2m_0}+m_0\Phi(\bs{x}),
\end{align}
$H_\text{INT}$ is the INT Hamiltonian given in \eqref{eq:HINT}, and $\Gamma$ is the red-shift factor.
We take a unit of $c=1$ in the following.

Using Eq.~\eqref{eq:H1}, if the particle is trapped in a harmonic
potential with the stiffness constant $k$,  the total Hamiltonian can
be written as follows \cite{Haustein2019}:
\begin{align}
  \hat H&=\hat m+\frac{\hat p^2}{2\hat m}+\hat m\,\hat\Phi(\hat
         x)+\frac{k}{2}\,\hat x^2
     \notag \\
  &=\sum_{j=0,1}\left[m_j+\frac{\hat p^2}{2m_j}+m_j\hat \Phi(\hat
    x)+\frac{m_j\omega_j^2}{2}\,\hat x^2\right]\otimes\ket{E_j}\bra{E_j}
  =\sum_{j=0,1} \hat H_j\otimes\ket{E_j}\bra{E_j},
  \label{eq:clock_Hamiltonian}
\end{align}
where  $\omega_j=(k/m_j)^{1/2}$.
Note that the internal energy level $E_0,~E_1$ are eigenvalues of $H_{\text{INT}}$, and they have nothing to do with an infinite number of energy eigenvalues of $H_{\text{CM}}$ since the INT and CM systems are orthogonal. To see that the particle system consists of two independent systems, CM and INT, explicitly, we can rewrite Eq.~\eqref{eq:clock_Hamiltonian} as
\begin{align}
\hat H=\sum_{j=0,1} \left\{\sum_{n=0}^\infty \left(m_j+\hbar\omega_j(n+1/2)\right) |n_j \rangle \langle n_j| \right\}\otimes|E_j \rangle \langle E_j|
\end{align}
where $n$ labels the energy eigenvalues of the CM system, and $|n_j\rangle$ is the $n$th eigenstate of the CM system for $\hat H_j$.

\begin{figure}[H]
  \centering
  \includegraphics[width=0.6\linewidth,clip]{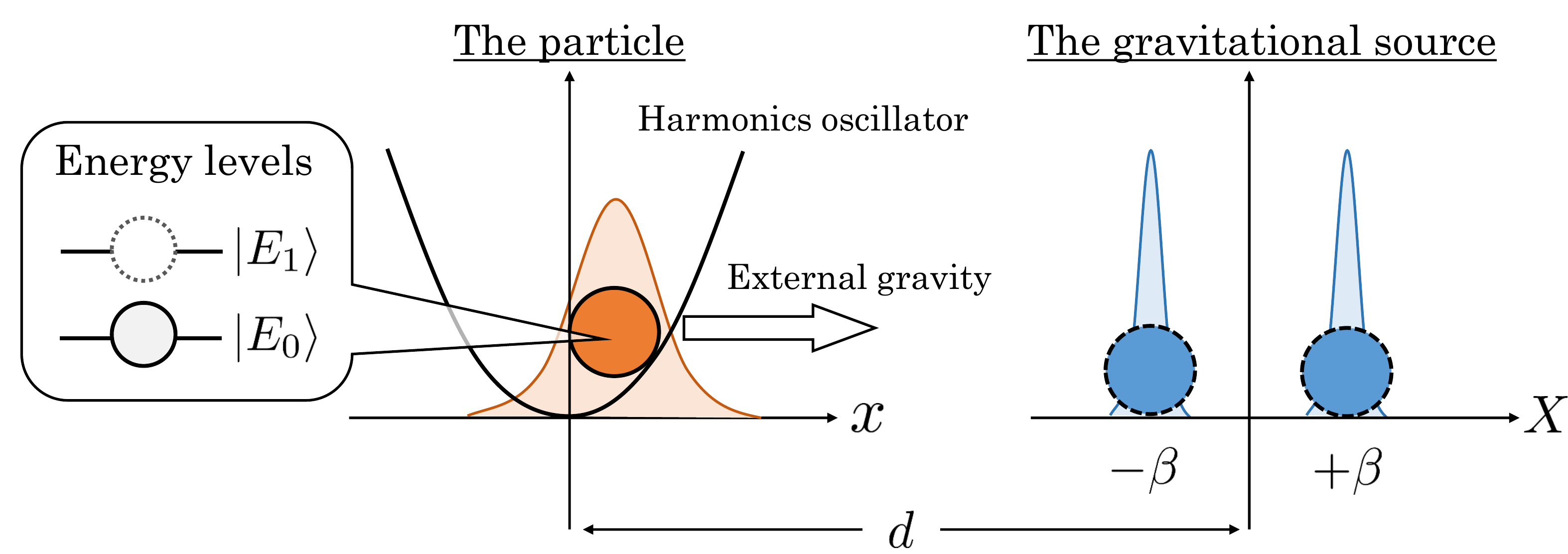}
  \caption{Setup for the proposed method. The gravitational source is placed
    at $X=\pm\beta$ as a quantum mechanically superposed state (cat
    state), and a trapped particle with internal energy levels
    interacts with it.}
  \label{fig:setup}
\end{figure}

Figure~\ref{fig:setup} displays the setup for the external
gravitational potential $\hat\Phi$, 
which form depends on whether gravity creates the quantum entanglement or not.
The gravitational
source mass is placed at $d+\hat X$. 
In Sec.~\ref{sec:Main} we will consider that the gravitational source is spacially superposed at $X=\pm\beta$. 
\noindent

We first focus on the case when gravity produces the quantum entanglement between the particle and the gravitational source systems. We will consider a specific form of the Newtonian potential which contains both the operators of the particle and the gravitational source systems, which we call quantized Newtonian gravity (QG) throughout the text. Note that quantized Newtonian gravity does not immediately imply the Newtonian limit of quantum gravity theory, but we adopt it as a particular gravitational interaction that produces entanglement. The QG potential is given as

\begin{align}
    \hat\Phi&=\frac{-GM}{d+\hat X-\hat x} 
    \approx
      -\frac{GM}{d}\left\{1+\frac{\hat{x}-\hat{X}}{d}+\left(\frac{\hat{x}-\hat{X}}{d}\right)^2\right\}
  \label{eq:QGpotential} \\
    &= -A(\hat X)-B(\hat X)\,\hat x-C\, \hat x^2,
      \label{eq:potential_expansion}
\end{align}
where
\begin{align}
    A(\hat X):=\frac{GM}{d}\left(1-\frac{\hat X}{d}+\frac{\hat X^2}{d^2}\right),\quad
    B(\hat X):=\frac{GM}{d^2}\left(1-\frac{2\hat X}{d}\right),\quad
    C:=\frac{GM}{d^3}.
    \label{eq:AB}
\end{align}
In the second equality of Eq.~\eqref{eq:QGpotential}, we performed Taylor expansion for $d\gg |\langle\hat X^n\rangle|,|\langle\hat x^n\rangle|~(n=0,1,2,\cdots)$.
Therefore, the Hamiltonian is
\begin{equation}
  \hat H_\text{QG}=\sum_{j=0,1}\left[\mathcal{E}_j(\hat X)+\frac{\hat p^2}{2m_j}
  +\frac{m_j\omega_j^2}{2}\left(\hat x-\Delta_j(\hat X)\right)^2\right]\ket{E_j}\bra{E_j},
\label{eq:HQG}
\end{equation}
where 
\begin{align}
    \mathcal{E}_j(\hat X):=m_j\left(1-A(\hat X)-\frac{B^2(\hat
    X)}{2\omega_j^2}\right),\quad
    \Delta_j(\hat X):=B(\hat X)/\omega_j^2.
\end{align}
Herein, we redefined a shifted angular frequency as $\omega_j^2-C\to\omega_j^2$.  $\mathcal{E}_j(\hat X)$ represents the offset of the
total energy of the CM system with the $j$th INT state, and
$\Delta_j(\hat X)$ represents the shift in the harmonic potential due
to the weak external
gravitational field.

\begin{figure}[H]
  \centering
  \includegraphics[width=0.4\linewidth,clip]{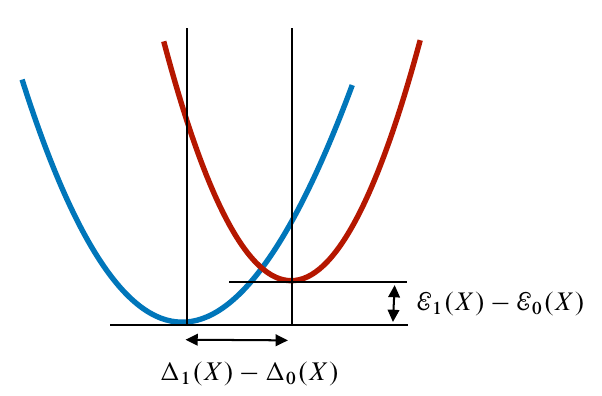}
  \caption{Schematic representation of the potential for CM
    degrees of freedom. Blue curve: the potential of
    $\hat H_0$. Red curve: the potential of
    $\hat H_1$. Shifts in the symmetry axis and the vertex height of
    the potential are caused by the differences
    $\Delta_1(X)-\Delta_0(X)$ and
    $\mathcal{E}_1(X)-\mathcal{\hat E}_0(X)$. The difference in
    the vertex height depends on the location of the gravitational
    source $X$.  }
  \label{fig:pot}
\end{figure}

The information of source position $\hat X$ is included in
$\mathcal{E}_j(\hat X)$ and $\Delta_j(\hat X)$, which indicates the
entanglement between the CM system $\hat x$ and the gravitational
source (S) system $\hat X$. Moreover, since the INT system also
couples with the CM system and  S system as well, the total system
is in a tripartite entangled state. In particular, the entanglement
between the INT system and S system is uniquely induced by the
gravitational coupling $\hat m\,\hat{\Phi}(\hat{x},\hat{X})$; No other
quantum interaction can create this entanglement since they do not
couple to the mass $\hat m$ due to the weak equivalence principle.
Further discussion will be given in Sec.~\ref{sec:Discussion}.
Figure \ref{fig:pot} displays a schematic representation of the
potential of  Hamiltonian $\hat H_j$. Shifts
of the vertex height and the symmetry axis of the potential depend on
INT states and this behavior is caused by the coupling between
the CM system and INT system via the weak external gravitational
field. Section \ref{sec:Main} will discuss
 that these shifts of the potential result in collapse
and revival of  interferometric visibility, which can be interpreted
as the universal decoherence of the INT state via external gravitational
field \cite{Zych2016,Pikovski2015}.
In the semiclassical treatment of gravity, hereafter referred to as classical
gravity (CG), $\hat\Phi$ is replaced by the expectation
value concerning the state of the gravitational source
$\ave{\hat\Phi}_S=\br{\varphi_S}\hat\Phi\kt{\varphi_S}$
which does not create quantum entanglement between the particle and the gravitational source systems ,unlike the QG case. Therefore, for $d\gg |\langle\hat X^n\rangle_S|,|\langle\hat x^n\rangle|~(n=0,1,2,\cdots)$
, the gravitational potential is approximately
given by
\begin{align}
    \hat\Phi
    &=\left\langle \frac{-GM}{d+\hat X-\hat x}\right\rangle_S
    \approx -\frac{GM}{d}\left\{1+\frac{\hat x-\ave{\hat
      X}_S}{d}+\frac{\hat x^2-2\ave{\hat X}_S\,\hat{x}+\ave{\hat
      X^2}_S}{d^2}\right\} \label{eq:CGpotential} \\
    &=-A-B\,\hat x-C\,\hat x^2,
\end{align}
where
\begin{align}
  A:=\frac{GM}{d}\left(1-\frac{\ave{\hat X}_S}{d}
  +\frac{\ave{\hat X^2}_S}{d^2}\right),\quad
  B:=\frac{GM}{d^2}\left(1-\frac{2\ave{\hat X}_S}{d}\right),\quad
  C:=\frac{GM}{d^3}.
\end{align}
Therefore, the corresponding
Hamiltonian becomes
\begin{equation}
  \hat H_\text{CG}=\sum_{j=0,1}\left[\mathcal{E}_j+\frac{\hat p^2}{2m_j}
  +\frac{m_j\omega_j^2}{2}\left(\hat x-\Delta_j\right)^2\right]\ket{E_j}\bra{E_j},
\end{equation}
where 
\begin{align}
    \mathcal{E}_j=m_j\left(1-A-\frac{B^2}{2\omega_j^2}\right),\quad
    \Delta_j=B/\omega_j^2.
    \label{eq:E_CG}
\end{align}
In the CG case, CM and INT can entangle through relativistic and
classical gravitational couplings, while INT and S cannot entangle.

Next,  we introduce the annihilation operator for the CM
system for later use:
\begin{equation}
  \hat a_{j,\hat X}=\sqrt{\frac{m_j\omega_j}{2}}
  \left(\hat x+i\frac{\hat p}{m_j\omega_j}-\Delta_j(\hat X)\right).
  \label{eq:annihilation_op}
\end{equation}
Therefore, the Hamiltonian for the QG case becomes
\begin{equation}
  \hat H_\text{QG}(\hat X)=\sum_{j=0,1}
  \left[\mathcal{E}_j(\hat X)+\omega_j\left(\hat a_{j,\hat X}{}^{\!\!\!\!\!\!\!\!\dag}\,\,\,\,\,\hat
      a_{j,\hat X}+\frac{1}{2}\right)\right]\ket{E_j}\bra{E_j}.
  \label{eq:HQG2}
\end{equation}
The Hamiltonian of the CG case is obtained by  replacing the functions of
$\hat X$ with the expectation value for the S state, $f(\hat X)\to\langle f(\hat{X})\rangle_S$, in
Eq. \eqref{eq:HQG2}.

\section{Evolution of the particle state and Visibility}
\label{sec:Main}
\begin{figure}[htbp]
  \centering
  \includegraphics[width=0.5\linewidth,clip]{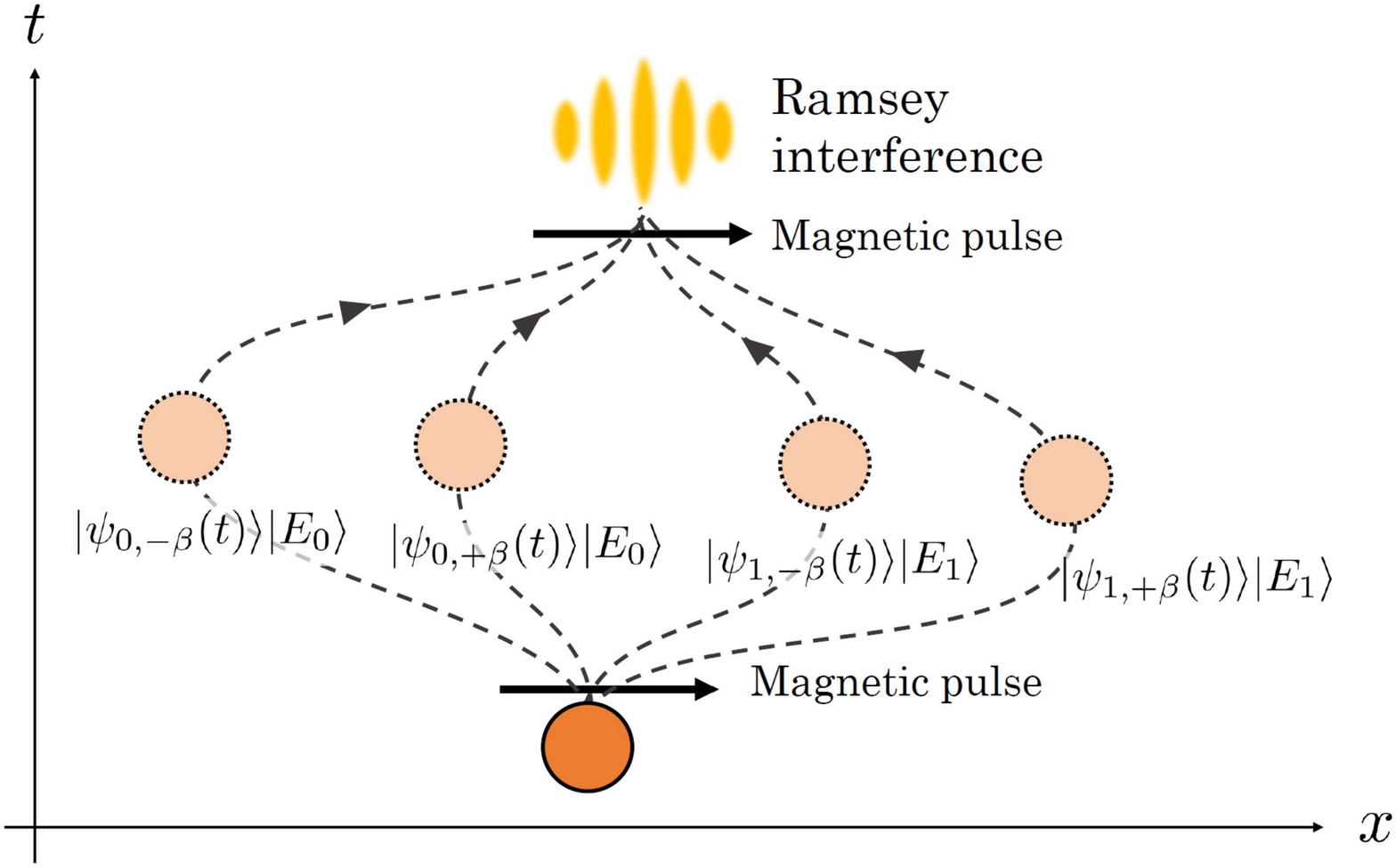}
  \caption{Schematic picture of  Ramsey
    interferometry in the QG case. Four different worldline branches
  are involved in  interference visibility.}
  \label{fig:interf_QG}
\end{figure}

In this section, we evaluate the interference visibility.
We assume that the gravitational source state is in superposition of two Gaussian states (See Fig. \ref{fig:setup}),
\begin{equation}
  \ket{\varphi_S}=\frac{1}{\sqrt{N}}\left(\ket{\varphi_{-\beta}}+\ket{\varphi_{+\beta}}\right),
  \quad
  \varphi_{\pm\beta}(X)=\bra{X}\ket{\varphi_{\pm\beta}}=
  \left(\frac{1}{\pi\sigma^2}\right)^{1/4}e^{-(X\mp\beta)^2/(2\sigma^2)},
  \quad
  N=2(1+e^{-\beta^2/\sigma^2}),
  \label{eq:cat-state}
\end{equation}
where $\beta$ is the separation of the cat state and $\sigma$ is the width
of  each Gaussian state.  The evolution operator associated with
the $j$th INT state can be expressed as
\begin{equation}
 e^{-i\mathcal{E}_j(\hat X)t}\,e^{-i\omega_j(\hat
   a_{j,\hat X}{}^{\!\!\!\!\!\!\!\!\dag}\,\,\,\,\hat a_{j,\hat X}+1/2)t}
 =e^{-i\mathcal{E}_j(\hat X)t}\,\hat U_{j,\hat
    X}(t),
\end{equation}
where $\hat U_{j,\hat X}(t)=e^{-i\hat h_j(\hat X)t}$ is the evolution
operator with the harmonic oscillator Hamiltonian
$\hat h_j(\hat X):=\omega_j(\hat a_{j,\hat X}{}^{\!\!\!\!\!\!\!\!\dag}\,\,\,\,\hat a_{j,\hat
  X}+1/2)$.
As the initial state of the CM system, we prepare the ground
state of $\hat h_0(X=0)$ as
\begin{align}
    \psi_{\text{ini}}(x)=\left(\frac{a_0}{\pi}\right)^{1/4}\exp\left[
    -\frac{a_0}{2}(x-\Delta_0(0))^2\right],\quad a_0=m_0\,\omega_0.
\end{align}
Therefore, the time evolution of the total state associated with the $j$th
INT state becomes
\begin{equation}
  \ket{\Psi_j(t)}=e^{-i\mathcal{E}_j(\hat X)t}\,\hat
  U_{j,\hat X}(t)\ket{\psi_{\text{ini}}}\otimes\ket{\varphi_S}
  =\int dX\, \varphi_S(X)\,
  e^{-i\mathcal{E}_j(X)t}\ket{\psi_{j,X}(t)}\otimes\ket{X},
  \label{eq:time_evolution}
\end{equation}
Here, we defined
$\ket{\psi_{j,X}(t)}:=\hat U_{j,X}(t)\ket{\psi_{\text{ini}}}$. 
Introducing the propagator of the harmonics oscillator, we obtain
\begin{align}
    &K_{j,X}(x,t;y,0)
    :=\bra{x}\hat U_{j,X}(t)\ket{y}\notag \\
    &=\sqrt{\frac{a_j}{2\pi
    i\sin(\omega_jt)}}\exp\left[\frac{ia_j}{2\sin(\omega_jt)}
    \biggl\{((x-\Delta_j(X))^2+(y-\Delta_j(X))^2)\cos(\omega_j
    t)-2(x-\Delta_j(X))(y-\Delta_j(X))\biggr\}\right],
\end{align}
and the wave function of the CM
state can be explicitly calculated as \cite{Gersch1992} 
\begin{align}
    &\psi_{j,X}(x,t)
    =\bra{x}\hat U_{j,X}(t)\ket{\psi_{\text{ini}}}
    =\int dy\bra{x}\hat U_{j,X}(t)\ket{y}\bra{y}\ket{\psi_{\text{ini}}}
    =\int dy\, K_{j,X}(x,t;y,0)\psi_{\text{ini}}(y) \notag\\
    &=\left(\frac{a_0}{\pi}\right)^{1/4}\sqrt{\frac{a_j}{a_j\cos(\omega_jt)
    +ia_0\sin(\omega_jt)}}
    \,\exp\left[-\frac{a_j^2}{2a_0}\frac{R_{j,X}(t,x)+iI_{j,X}(t,x)}{1+(a_j^2/a_0^2-1)\cos^2(\omega_j
    t)}\right],
    \label{eq:time_evolved_psi}
\end{align}
where $a_j:=m_j\omega_j$ and 
\begin{align}
  &R_{j,X}(t,x):=\biggl[(x-\Delta_j(X))-(\Delta_0(X)-\Delta_j(X))\cos(\omega_j
    t)\biggr]^2,\\
  &I_{j,X}(t,x):=-\sin(\omega_jt)\left[\frac{a_0}{a_j}
    \biggl\{\left((x-\Delta_j(X))^2+(\Delta_0(X)-\Delta_j(X))^2\right)\cos(\omega_j
    t)-2(x-\Delta_j(X))(\Delta_0(X)-\Delta_j(X))\biggr\}    \right.\notag\\
    &\left.\hspace{250pt}-\frac{a_j}{a_0}(x-\Delta_j(X))^2\cos(\omega_jt)\right].
\end{align}
To obtain the transition probability Eq.~\eqref{eq:prob}, we should
evaluate 
\begin{align}
  \bra{\Psi_0(t)}\ket{\Psi_1(t)}
  &=\int dX\, |\varphi_S(X)|^2 \, e^{-i(\mathcal{E}_1(X)-\mathcal{E}_0(X))t}
    \bra{\psi_{0,X}(t)}
    \ket{\psi_{1,X}(t)}.
   \label{eq:V1}
\end{align}
The source state is assumed to be prepared as Eq.~\eqref{eq:cat-state} to reveal quantum superposition on gravity,
and has no dynamics for simplicity of treatment. Furthermore, the S system is regarded as a
two-level state with $X=\pm\beta$. Therefore,
Eq.~\eqref{eq:time_evolution} reduces to the following expression,
\begin{align}
  \ket{\Psi_j(t)}\approx \frac{1}{\sqrt{N}}
  \sum_{s=\pm\beta}e^{-i\mathcal{E}_j(s)t}\ket{\psi_{j,s}(t)}\otimes\ket{\varphi_{s}}.
  \label{eq:state-approx}
\end{align}
Herein, we made the assumption that $\psi_{j,X}(t,x)$ and $e^{-i\mathcal{E}_j(X)t}$
in Eq. \eqref{eq:time_evolution} 
do not change rapidly with respect to $X$ within the width  $\sigma$
of the Gaussian function in
$\varphi_S(X)$. Each assumption requires the following condition
respectively,
\begin{align}
    a_j \left(\frac{GM}{\omega^2 d^3}\right)^2
    \ll \frac{1}{\sigma^2},\quad 
    \frac{Gm_jM}{d^3}t \ll \frac{1}{\sigma^2}.
    \label{eq:t_condition}
\end{align}
The former inequality is satisfied when the Gaussian dispersion of the
CM system and the source system are in the same order, and the
displacement of the CM state is relatively small compared to 
$d$: $\frac{GM}{\omega^2 d^3}\sim \Delta_j/d\ll1$. The latter
inequality gives the upper bound of the valid time range $t$ for the
form of the state expressed in Eq. \eqref{eq:state-approx}.  Then
Eq.~\eqref{eq:V1} is approximately given as a summation of four terms,
\begin{align}
    \bra{\Psi_0(t)}\ket{\Psi_1(t)}
    \approx\frac{1}{N}\!\!\sum_{s_1,s_2=\pm\beta}\!\!
  \bra{\varphi_{s_1}}\ket{\varphi_{s_2}}e^{-i(\mathcal{E}_1(s_2)
  -\mathcal{E}_0(s_1))t}
    \bra{\psi_{0,s_1}(t)}
    \ket{\psi_{1,s_2}(t)}.
    \label{eq:V1_qubit}
\end{align}
Figure~\ref{fig:interf_QG} depicts a schematic picture of the Ramsey
interferometry under quantum gravity (QG). Since the CM state depends not only on the
INT energy level $E_i$ but also the gravitational source position
$X=\pm\beta$, its time evolution is given by a superposition of four
different worldline branches (Additional details regarding the states
involved in the interference visibility are given in Appendix \ref{appendix}).

For the  CG case, the Ramsey interferometry is derived 
analogously  by replacing  the functions of $\hat{X}$ with
the expectation value of $f(\hat X)\to\langle f(\hat
X)\rangle_S$. Then, Eq.~\eqref{eq:V1} can be expressed as
\begin{align}
    \bra{\Psi_0(t)}\ket{\Psi_1(t)}
    =e^{-i(\mathcal{E}_1-\mathcal{E}_0)t}\bra{\psi_0(t)}\ket{\psi_1(t)},
    \label{eq:V1_qubit_CG}
\end{align}
where $\mathcal{E}_j$ is as per in Eq.~\eqref{eq:E_CG}, and
$\ket{\psi_j(t)}:=e^{-i\hat{h}_j t}\ket{\psi_{\text{ini}}}$ is the time
evolved CM state with the harmonics oscillator Hamiltonian for CG
$\hat{h}_j:=\langle \hat{h}_j(\hat X)\rangle_S$. The Ramsey
interferometry under CG is depicted in Fig.~\ref{fig:interf_CG}. We
see that the time evolution of the CM system is given by a
superposition of two worldline branches concerning the INT energy levels
$E_0$ or $E_1$.

\begin{figure}[H]
  \centering
    \includegraphics[width=0.45\linewidth,clip]{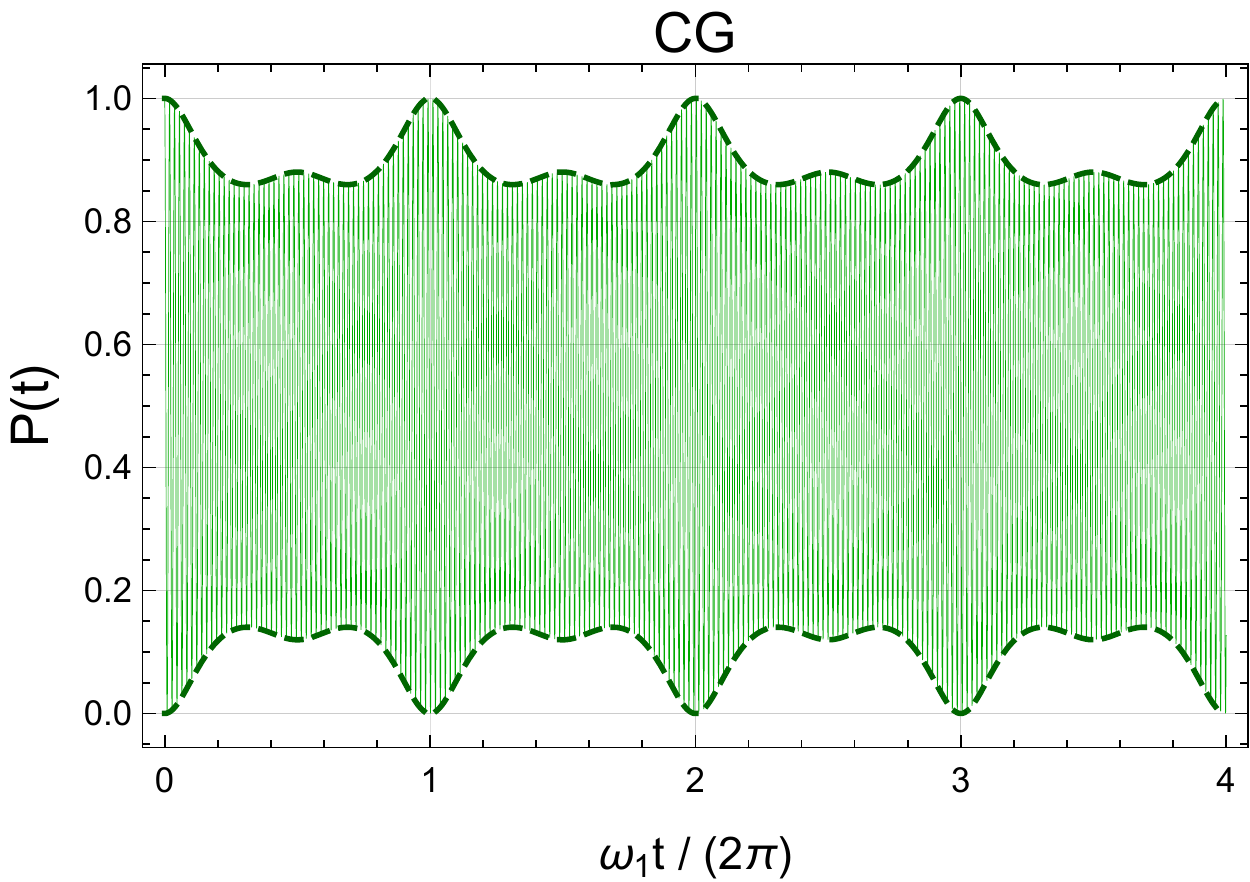}
    \includegraphics[width=0.45\linewidth,clip]{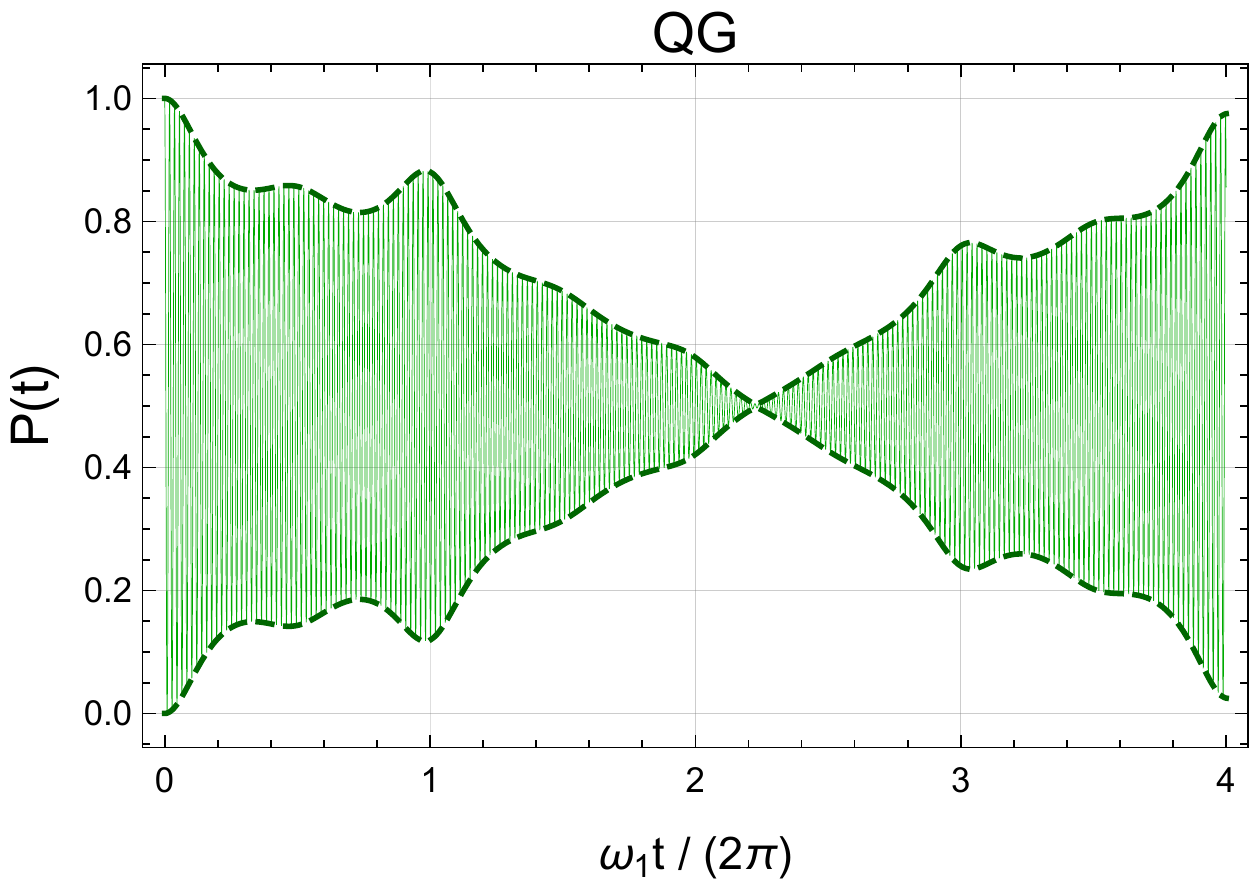}
    \caption{Probability $P(t)$ of the CG case (left panel) and QG
      case (right panel). Solid line: time dependence of the
      probability $P(t)$. Dashed line: the envelope of $P(t)$, which
      is $\frac{1}{2}(1+|\mathcal{V}(t)|)$. }
 \label{fig:probability}
\end{figure}

\begin{figure}[H]
  \centering
  \includegraphics[width=0.5\linewidth,clip]{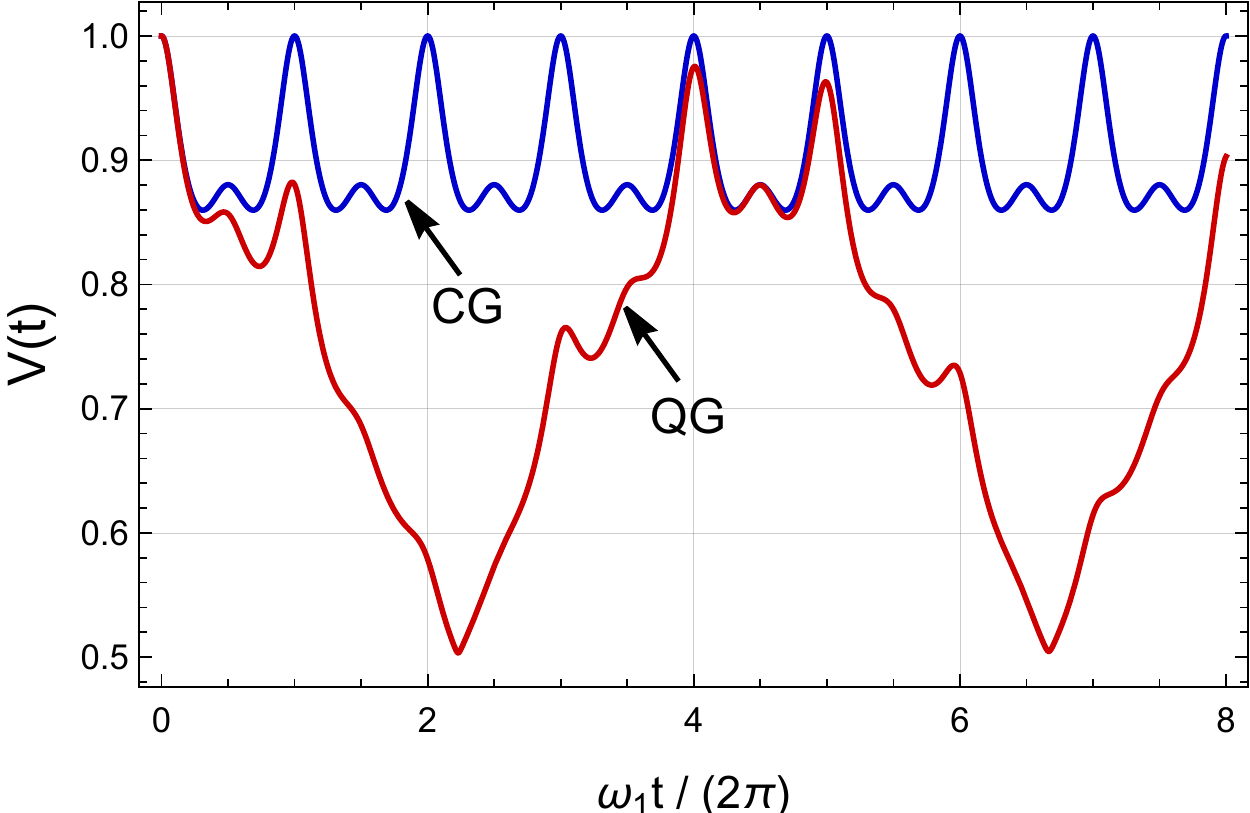}
  \caption{Visibility for the CG case  (blue line) and the
    QG case (red line). Owing to the decoherence effect induced by
      the spread of the source state, the visibility of QG does
        not come back to unity and does not
      show revival behavior.}
  \label{fig:visibility_CGQG}
\end{figure}

Although it is possible to derive an explicit but complicated formula
of Eq.~\eqref{eq:V1_qubit} using a straightforward calculation of the 
Gaussian integral, we  only present the plotting results in 
subsequent figures Figs. \ref{fig:probability},
\ref{fig:visibility_CGQG}, \ref{fig:lognega}, and \ref{fig:probability_Coulomb}. Instead, in the following, we derive an approximate
analytical form of Eq. \eqref{eq:V1} by  focusing only up to the leading
term of the Taylor expansion of the gravitational potential expressed in
Eq.~\eqref{eq:potential_expansion}, which helps our qualitative
understanding of gravity-induced decoherence.  If we neglect the
subleading terms of the Taylor expansion parameters $x/d$ and $X/d$ in \eqref{eq:potential_expansion},
Eq.~\eqref{eq:AB} can be expressed as
\begin{align}
    A(\hat{X})\to\frac{GM}{d}\left(1-\frac{\hat{X}}{d}\right),\quad
    B(\hat{X})\to \frac{GM}{d^2},\quad
    C\to 1.
    \label{eq:AB_firstorder}
\end{align}
Herein, the $\hat{X}$ dependence only appears in
$\mathcal{E}_j(\hat{X})$, and the coupling of the CM system operator
$\hat{x}$ and the S system operator $\hat{X}$ disappears
as $\Delta_j(\hat X)\to\Delta_j$. Although the BMV
proposal \cite{Bose2017,Marletto2017} focuses on the entanglement
between the CM system and  S system, this study
 aims to investigate the entanglement between the INT system
and S system, which is uniquely induced by gravitational
coupling due to the weak equivalence principle. (Further details will be revisited in Sec.~\ref{sec:Discussion}.) Therefore, this
approximation is enough to capture the crucial effect of the quantumness
of gravity in our setup of the Ramsey interferometry.

Since $\hat h_j$ does not depend on $\hat X$ for now, 
the $X$ integration in Eq. \eqref{eq:V1} can be simplified to
\begin{align}
    \bra{\Psi_0(t)}\ket{\Psi_1(t)}
    &=\bra{\psi_{0}(t)}\ket{\psi_{1}(t)}
      \times\int dX\, |\varphi_S(X)|^2\,
      e^{-i(\mathcal{E}_1(X)-\mathcal{E}_0(X))t}.
    \label{eq:V2}
\end{align}
Evaluating the inner product of CM states
$\bra{\psi_{0}(t)}\ket{\psi_{1}(t)}$ with \eqref{eq:AB_firstorder}, we obtain
\begin{align}
    \bra{\psi_{0}(t)}\ket{\psi_{1}(t)}
    =\int dx\,\psi_{0}^*(t,x)\psi_{1}(t,x)
    =:
    |\mathcal{V}_C(t)|\,e^{i\Theta_C(t)},
    \label{eq:visibility}
\end{align}
where
\begin{align}
    |\mathcal{V}_C(t)|
  &=\left(\frac{4a_0\,a_1}{4a_0\,a_1\cos^2(\omega_1t)+(a_0+a_1)^2
    \sin^2(\omega_1 t)}\right)^{1/4}
    \exp\left[
    -\frac{2a_0\,a_1^2(\Delta_0-\Delta_1)^2
    \sin^2(\omega_1t/2)}{a_0^2+a_1^2+(a_0^2-a_1^2)\cos(\omega_1t)}
    \right]
    \label{eq:VC},\\
    \Theta_C(t)
  &=-\frac{a_0\, a_1^2\left(\Delta_0-\Delta_1\right)^2
    \sin(\omega_1 t)}{a_0^2+a_1^2+(a_0^2-a_1^2)\cos(\omega_1 t)}
    -\frac{1}{2}\arg\left[e^{-i\omega_0 t}\left(2a_0/a_1\cos(\omega_1
    t)
    +i\left(1+(a_0/a_1)^2\right)\sin(\omega_1 t)\right)\right].
    \label{eq:thetaC}
\end{align}
It should be noted that $\mathcal{V}_C(t)$ is a $2\pi/\omega_1$
periodic function in time $t$ that reflects the oscillation of the CM state in
the harmonic oscillator potential. The $X$ integration in Eq.~\eqref{eq:V2} reads to
\begin{align}
    \int dX\, |\varphi_S(X)|^2\,  e^{-i(\mathcal{E}_1(X)-\mathcal{E}_0(X))t}
    =|\mathcal{V}_Q(t)|\,e^{i\Theta_{\mathcal{E}}(t)},
    \label{eq:X_integration}
\end{align}
where 
\begin{align}
  |\mathcal{V}_Q(t)|&=\frac{2}{N}
                      \exp\left[-\left(\frac{GME}{2d}
                      \frac{\sigma}{d}\,t\right)^2\right]\left|
                      \cos\left(\frac{GME}{d}\frac{\beta}{d}\,t\right)
                      +e^{-\beta^2/\sigma^2}\right|,
    \label{eq:VQ}\\
    \Theta_{\mathcal{E}}(t)&=-(\mathcal{E}_1(0)-\mathcal{E}_0(0))\,t.
    \label{eq:thetaE}
\end{align}
Finally, when we treat up to the leading term of $x/d,~X/d$, the
analytic expression of Eq.~\eqref{eq:V1} is given by
\begin{align}
  \bra{\Psi_0(t)}\ket{\Psi_1(t)}=|\mathcal{V}_C(t)
  \mathcal{V}_Q(t)|\,e^{i(\Theta_C(t)+\Theta_{\mathcal{E}}(t))}
  \qquad\text{(QG case)},
    \label{eq:V_QG}
\end{align}
where the time-dependent functions $|\mathcal{V}_C|, \Theta_C,
|\mathcal{V}_Q|, \Theta_\mathcal{E}$ are given in Eqs. \eqref{eq:VC},
\eqref{eq:thetaC}, \eqref{eq:VQ}, and \eqref{eq:thetaE}, respectively.
For the CG case, the  treatment can be applied  by replacing
$\hat X\to \langle X\rangle=0$, which reduces the $X$ integration in
Eq.~\eqref{eq:X_integration} to 1, and we get the final
expression for the inner product as
\begin{align}
  \bra{\Psi_0(t)}\ket{\Psi_1(t)}=|\mathcal{V}_C(t)|\,e^{i(\Theta_C(t)
  +\Theta_{\mathcal{E}}(t))}
  \qquad\text{(CG case)}.
    \label{eq:V_CG}
\end{align}

Let us discuss some properties of interference
for different treatments of gravity based on Eqs. \eqref{eq:V_QG} and
\eqref{eq:V_CG} in the following.

First, when there is no external gravitational
source ($G=0$ and $\Delta_0=\Delta_1=0$), the probability oscillates
with a period determined by the energy gap of the INT system
$\mathcal{E}_1-\mathcal{E}_0=E$. The envelope of the oscillation is determined by visibility $|\mathcal{V}_C(t)|$, whose time period $\pi/\omega_1$ reflects the 
evolution of the squeezed state, as seen Eq. \eqref{eq:thetaC} (and
also in Appendix \ref{appendix}). The decoherence arises from  the entanglement between the CM system and
 INT systems induced by the kinetic term. In other words, a particle with a
different energy level evolves along a different branch of world line with a
different proper times due to the special relativistic redshift, as
seen in Fig.~\ref{fig:interf_CG}, which results in  universal decoherence
\cite{Pikovski2015}.

For the CG case,the
visibility factor $|\mathcal{V}_C(t)|$ contains two typical periods: (i) $\pi/\omega_1$, which corresponds to the period of
squeezed state, and (ii) $2\pi/\omega_1$, which corresponds to the
period of coherent state originated from the interaction of the CM state
with the external CG \cite{Haustein2019} (see also Appendix
\ref{appendix}). Each periodic decoherence behavior in CG occurs due
to the entanglement between the CM and the INT systems induced by the
kinetic term, and the semiclassical gravitational interaction; the
later refers to universal decoherence caused by a gravitational
redshift \cite{Pikovski2015}.  In the left panel of
Fig.~\ref{fig:probability}, we show the time
dependence of the probability $P(t)$ under CG with a green solid line, which is obtained by
evaluating Eq. \eqref{eq:V1_qubit_CG}. It exhibits oscillation with a
period $\approx 2\pi/(\mathcal{E}_1-\mathcal{E}_0)$, which roughly
corresponds to the energy gap of the INT system with relativistic and
gravitational corrections. The dashed lines denote its envelope
$(1+|\mathcal{V}_C(t)|)/2$, and parameters are chosen as
$(m_0^3/k)^{1/2}=10,~(m_0/m_1)^{1/2}=0.5,~Gm_0M/(kd^3)=0.015,~m_0^{1/4}d=10,~\beta/d=0.01,~\sigma/d=0.001$.
The visibility again exhibits revival behavior with the
$2\pi/\omega_1$ period, thereby reflecting semi-classical gravity
induced entanglement.

Next, we conduct an in-depth evaluation of the visibility of the QG case in detail as the
main result of this study and clarify the effect of the quantum
superposition of the gravitational source on the visibility of the
Ramsey interferometry. As an effect of the quantumness of gravity, the
visibility additionally contains $|\mathcal{V}_Q(t)|$. Furthermore,  a
modulation with a longer period appears in the visibility due to 
the following  reasons: (i) the exponential factor in
$|\mathcal{V}_Q(t)|$ of Eq. \eqref{eq:VQ} causes the temporal decay of
coherence. It contains the gravitational coupling
between the source mass and the INT energy gap and vanishes for
$\sigma\to0$. This indicates the entanglement induced by the
quantum gravitational interaction between the INT system and the
Gaussian dispersion of the S system. Since its
decoherence time scale is given by
$t\sim \left(\frac{GME}{2d}\frac{\sigma}{d}\right)^{-1}$, the system
decoheres more rapidly for larger $\sigma$.  (ii) The product of
two kinds of periodic functions, namely the cosine function in
$\mathcal{V}_Q(t)$ and the function $\mathcal{V}_C(t)$ with a period
$2\pi/\omega_1$ causes a beat in the visibility. The cosine
function in $\mathcal{V}_Q(t)$ comes from the gravitational coupling
between the source mass and the INT energy gap, and it vanishes for
$\beta\to0$. This indicates the entanglement induced by the quantum
gravitational interaction between the INT system and the distant cat
state. Since its decoherence time scale is given by
$t\sim \left(\frac{GME}{d}\frac{\beta}{d}\right)^{-1}$, the system
decoheres more rapidly for larger $\beta$.

Decoherence effect (i) is caused by the Gaussian spread of the source
mass state, and causes nonrevival behavior of the visibility. It is
induced by the entanglement between the INT state $\ket{E_j}$ and the Gaussian
state $\int dX e^{-X^2/2\sigma^2}\ket{X}$ with infinite
rank. Decoherence effect (ii) is induced by the entanglement between the INT state and
the distant cat state $\ket{\pm\beta}$ with rank 2. Therefore, the
combination of the INT state and the distant cat state subsystem
mentioned in (ii) is much easier to get entangled compared to
the the combination of the INT state $\ket{E_j}$ and the Gaussian
state $\int dX e^{-X^2/2\sigma^2}\ket{X}$ with infinite rank mentioned
in (i).  To summarize, the quantumness of gravity is reflected
in the visibility as a nonrevival behavior of visibility
(coherence), which is induced by the  entanglement between the INT
system and the S system as mentioned in Sec.~\ref{sec:internal}.

In the right panel of Fig.~\ref{fig:probability}, we showed the time dependence of the probability $P(t)$ for the QG case with a solid line, and its envelope
$(1+|\mathcal{V}_C(t)\mathcal{V}_Q(t)|)/2$ with a dotted line. The parameters
are set to
$(m_0^3/k)^{1/2}=10,~(m_0/m_1)^{1/2}=0.5,~Gm_0M/(kd^3)=0.015,~m_0^{1/4}d=10,~\beta/d=0.01,~\sigma/d=0.001$. The
figure displays the beat in the envelope of its oscillation, and
unlike the CG case, the visibility does not exhibit revival
behavior. It should be noted 
that the exponential decay in $\mathcal{V}_{QG}(t)$ obtained in
Eq. \eqref{eq:VQ} is not obvious here, since the exponential factor in
Eq. \eqref{eq:V1} is not dominant in the integral unless a time $t$
violates the condition expressed in Eq. \eqref{eq:t_condition}.
 
Finally, the behavior of visibility is explored by comparing the CG and QG
cases displayed in Fig.~\ref{fig:visibility_CGQG}. The blue and red
lines denote the CG
case and QG case respectively. The parameters are identical to those mentioned in
Fig.~\ref{fig:probability}.  As per the figure, the visibility of QG returns to
  nearly one after a period of $(GME\beta/d^2)^{-1}$, while its value decays
by $\exp\left(-(GME\sigma t/2d^2)^2\right)$, as estimated in
Eq. \eqref{eq:VQ}. To summarize, the quantumness of
gravity appears as a nonrevival behavior of interference
visibility. Stronger decoherence in QG compared to that in CG indicates the entanglement
sharing between the particle and S systems.

\section{Negativity of reduced bipartite states}
\label{sec:Negativity}
In this section, we will evaluate the entanglement  between
the CM, INT, and the gravitational source systems, which provide a
better understanding of the decoherence in the Ramsey interference
mentioned in Sec. \ref{sec:Main}.  The state of the total system is
given by
\begin{align}
    \ket{\Psi(t)}&=
    \frac{1}{\sqrt{2}}\sum_{j=0,1}e^{-i\mathcal{E}_j(\hat
                   X)t}\,\kt{E_j}\otimes
                   \kt{\psi_{j,\hat X}(t)}\otimes\kt{\varphi_{S}}.
    \label{eq:Psi}
\end{align}
To see the entanglement structure during the collapse and revival of
coherence in Fig.~\ref{fig:visibility_CGQG}, we assumed the same
approximation adopted in Sec. \ref{sec:Main} for
Eq. \eqref{eq:t_condition}. As preparation,  the CM states
and the source Gaussian state are rewritten as
\begin{align}
    \begin{bmatrix}
        \kt{C_0}\\ \kt{C_1}\\ \kt{C_2}\\ \kt{C_3}
    \end{bmatrix}
    =
    \begin{bmatrix}
        \kt{C_{J[0,0]}}\\ \kt{C_{J[0,1]}}\\ \kt{C_{J[1,0]}}\\ \kt{C_{J[1,1]}}
    \end{bmatrix}
    :=
    \begin{bmatrix}
        \kt{\psi_{0,-\beta}}\\ \kt{\psi_{0,+\beta}}\\
        \kt{\psi_{1,-\beta}}\\
        \kt{\psi_{1,+\beta}}
    \end{bmatrix}
    ,\quad
    \begin{bmatrix}
        \kt{S_0}\\ \kt{S_1}
    \end{bmatrix}
    :=
    \begin{bmatrix}
        \kt{\varphi_{-\beta}}\\ \kt{\varphi_{+\beta}}
    \end{bmatrix},
\end{align}
where we define a the function $J[j,k]:=2j+k=\{0,1,2,3\}$, and the offset of the
total energy $\mathcal{E}_{J[j,k]}:=\mathcal{E}_j((2k-1)\beta)$.
The state of the total system is expressed as
\begin{align}
  \ket{\Psi(t)}&=
                 \frac{1}{\sqrt{2N}}\sum_{j=0,1}\sum_{k=0,1}
                 e^{-i\mathcal{E}_{J[j,k]}t}
                 \,\kt{E_j}\otimes\kt{C_{J[j,k]}}\otimes\kt{S_k}.
\end{align}
Therefore,  the density matrix of the total system is obtained as
\begin{equation}
  \rho=\ket{\Psi}\bra{\Psi}
  = \frac{1}{2N}\sum_{j_1,j_2}\sum_{k_1,k_2}
  e^{-i(\mathcal{E}_{J[j_1,k_1]}-\mathcal{E}_{J[j_2,k_2]})t}
  \ket{E_{j_1}}\bra{E_{j_2}}\otimes
  \kt{C_{J[j_1,k_1]}}\br{C_{J[j_2,k_2]}}\otimes
  \ket{S_{k_1}}\bra{S_{k_2}}.
\end{equation}
We considered three different reduced bipartite states as follows:
\begin{align}
    &\rho_\text{INT:CM}:=\mathrm{Tr}_\text{S}\rho
    =\frac{1}{2N}\!\!
      \sum_{j_1j_2,k_1k_2}
      e^{-i(\mathcal{E}_{J[j_1,k_1]}-\mathcal{E}_{J[j_2,k_2]})t}
    \bra{S_{k_2}}\ket{S_{k_1}}
    \ket{E_{j_1}}\bra{E_{j_2}}\otimes
    \kt{C_{J[j_1,k_1]}}\br{C_{J[j_2,k_2]}},\\
    &\rho_\text{INT:S}:=\mathrm{Tr}_\text{CM}\rho
      =\frac{1}{2N}\!\!\sum_{j_1j_2,k_1k_2}
      e^{-i(\mathcal{E}_{J[j_1,k_1]}-\mathcal{E}_{J[j_2,k_2]})t}
    \langle{C_{J[j_2,k_2]}}\kt{C_{J[j_1,k_1]}}
    \ket{E_{j_1}}\bra{E_{j_2}}\otimes 
    \ket{S_{k_1}}\bra{S_{k_2}},\\
    &\rho_\text{CM:S}:=\mathrm{Tr}_\text{INT}\rho
    =\frac{1}{2N}
    \sum_{j,k_1k_2}e^{-i(\mathcal{E}_{J[j,k_1]}-\mathcal{E}_{J[j,k_2]})t}
    \ket{S_{k_1}}\bra{S_{k_2}}\otimes\kt{C_{J[j,k_1]}}\br{C_{J[j,k_2]}}.
\end{align}
To evaluate the negativity for the reduced states, we introduced an
orthonormal basis for the S state and the CM state. For the
source state, using the orthonormal basis
$\{\ket{s_k}\},~k=\{0,1\},~\bra{s_{k_1}}\ket{s_{k_2}}=\del_{k_1k_2}$,
we obtain
\begin{equation}
    \begin{bmatrix}
        \ket{S_0}\\
        \ket{S_1}
    \end{bmatrix}
    =\frac{1}{\sqrt{2}}
    \begin{bmatrix}
        \sqrt{1+v_S}&\sqrt{1-v_S}\\
        \sqrt{1+v_S}&-\sqrt{1-v_S}
    \end{bmatrix}
    \begin{bmatrix}
        \ket{s_0}\\
        \ket{s_1}
    \end{bmatrix}
    ,\quad v_{S}=\bra{S_0}\ket{S_1}=e^{-\beta^2/\sigma^2}.
\end{equation}
This relation is written using a $2\times 2$ matrix $S$ as
$\ket{S_{k}}=\sum_{\ell=0,1}S_{k\ell}\ket{s_{\ell}}$.  For the CM
state $\kt{C_{J[j,k]}}$, the orthonormal basis
$\{\ket{c_J}\},~J=\{0,1,2,3\},
\bra{c_{J_1}}\ket{c_{J_2}}=\del_{J_1J_2}$ is obtained using the
Gram-Schmidt orthonormalization as follows:
\begin{equation}
    \begin{bmatrix}
        \kt{C_0}\\ \kt{C_1}\\ \kt{C_2}\\ \kt{C_3}
    \end{bmatrix}
    =
    \begin{bmatrix}
        \sqrt{N_0}&0&0&0\\
        \bra{c_0}\psi_1\rangle&\sqrt{N_1}&0&0 \\
        \bra{c_0}\psi_2\rangle&\bra{c_1}\psi_2\rangle&\sqrt{N_2}&0\\
        \bra{c_0}\psi_3\rangle&\bra{c_1}\psi_3\rangle&\bra{c_2}\psi_3\rangle&\sqrt{N_3}
    \end{bmatrix}
    \begin{bmatrix}
        \kt{c_0}\\\kt{c_1}\\\kt{c_2}\\\kt{c_3}
    \end{bmatrix}
    ,\quad\quad
    N_J=1-\sum_{0\le K\le J-1}|\langle c_K\kt{\psi_J}|^2.
\end{equation}
Equivalently, the CM state can be expressed as
$\kt{\psi_{J}}=\sum_{K=0,1,2,3}\mathcal{U}_{JK}\kt{c_K}$ using a
$4\times 4$ matrix $\mathcal{U}_{JK}$.  Therefore, the reduced states are
\begin{align}
    &\rho_\text{INT:CM}
    =\frac{1}{2N}\!\!
      \sum_{j_1j_2,k_1k_2}\!\!
      e^{-i(\mathcal{E}_{J[j_1,k_1]}-\mathcal{E}_{J[j_2,k_2]})t}
    \left(\sum_{\ell}S^*_{k_1\ell}\,S_{k_2\ell}\right)
    \kt{E_{j_1}}\br{E_{j_2}}\otimes
    \sum_{K_1K_2}\mathcal{U}_{J[j_1,k_1]K_1}\,\mathcal{U}^*_{J[j_2,k_2]K_2}\kt{c_{K_1}}\br{c_{K_2}},
    \label{eq:state_INTCM}\\
    &\rho_\text{INT:S}
    =\frac{1}{2N}\!\!
    \sum_{j_1j_2,k_1k_2}\!\!
    e^{-i(\mathcal{E}_{J[j_1,k_1]}-\mathcal{E}_{J[j_2,k_2]})t}
    \left(\sum_K\mathcal{U}^*_{J[j_1,k_1]K}\,\mathcal{U}_{J[j_2,k_2]K}\right)
    \kt{E_{j_1}}\br{E_{j_2}}
    \otimes\sum_{\ell_1\ell_2}S_{k_1\ell_1}\,S^*_{k_2\ell_2}\kt{s_{\ell_1}}\br{s_{\ell_2}},
    \label{eq:state_INTS}\\
    &\rho_\text{CM:S}
    =\frac{1}{2N}\!\!
    \sum_{j,k_1k_2}\!e^{-i(\mathcal{E}_{J[j,k_1]}-\mathcal{E}_{J[j,k_2]})t}
    \sum_{\ell_1\ell_2}S_{k_1\ell_1}\,S_{k_2\ell_2}\kt{s_{\ell_1}}\br{s_{\ell_2}}\otimes
    \sum_{K_1K_2}\mathcal{U}_{J[j,k_1]K_1}\,\mathcal{U}^*_{J[j,k_2]K_2}\kt{c_{K_1}}\br{c_{K_2}}.
    \label{eq:state_CMS}
\end{align}
The entanglement negativity \cite{Vidal2002a} for a bipartite state is
obtained as $\mathcal{N}=\sum_{\lambda_i<0}|\lambda_i|$ where
$\lambda_i$ is the eigenvalue of the partially transposed density
matrix. Logarithmic negativity is defined by
\begin{equation}
  N_E:=\log_2(2\mathcal{N}+1),
  \label{eq:lognega}
\end{equation}
which quantifies the distillable number of Bell pairs.

\begin{figure}[H]
    \centering
    \includegraphics[width=0.45\linewidth,clip]{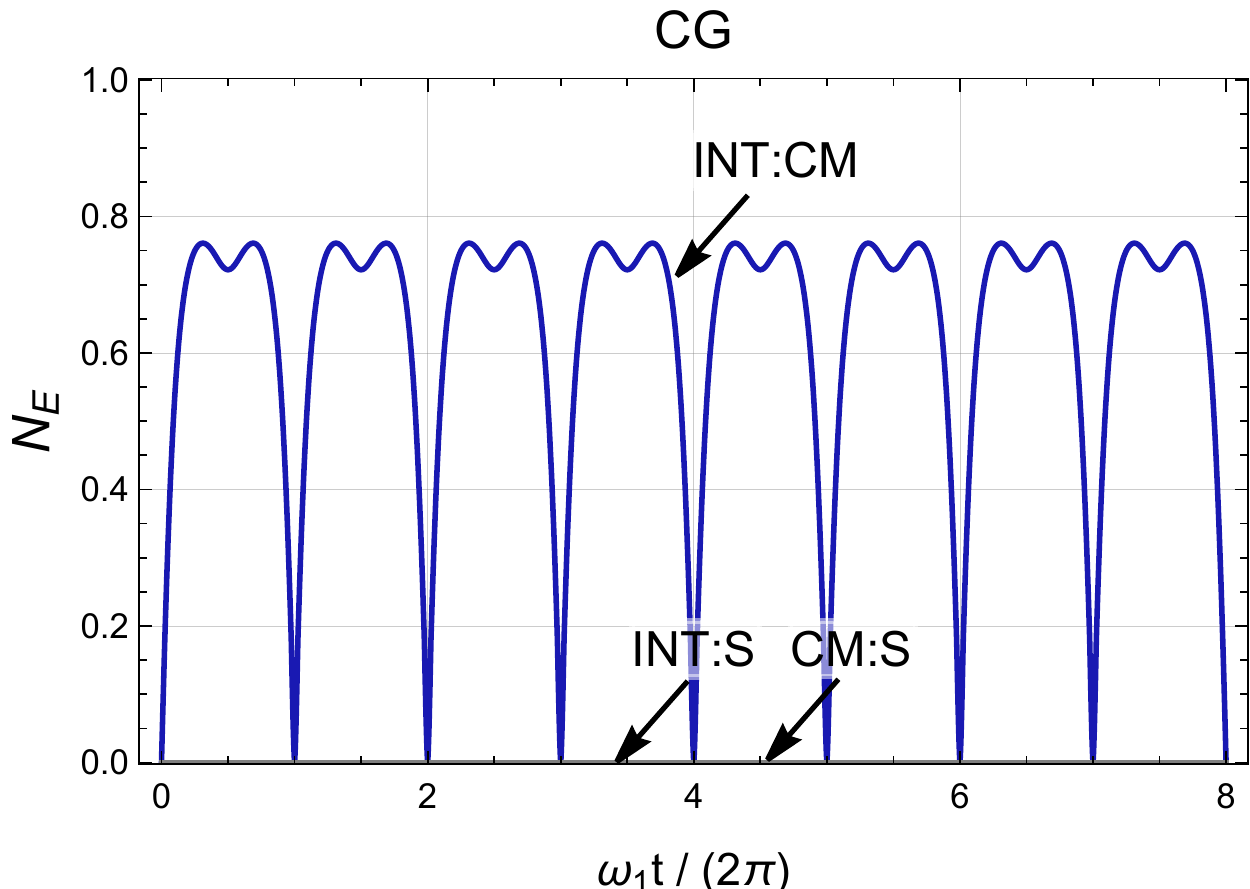}
    \hspace{0.5cm}
    \includegraphics[width=0.45\linewidth,clip]{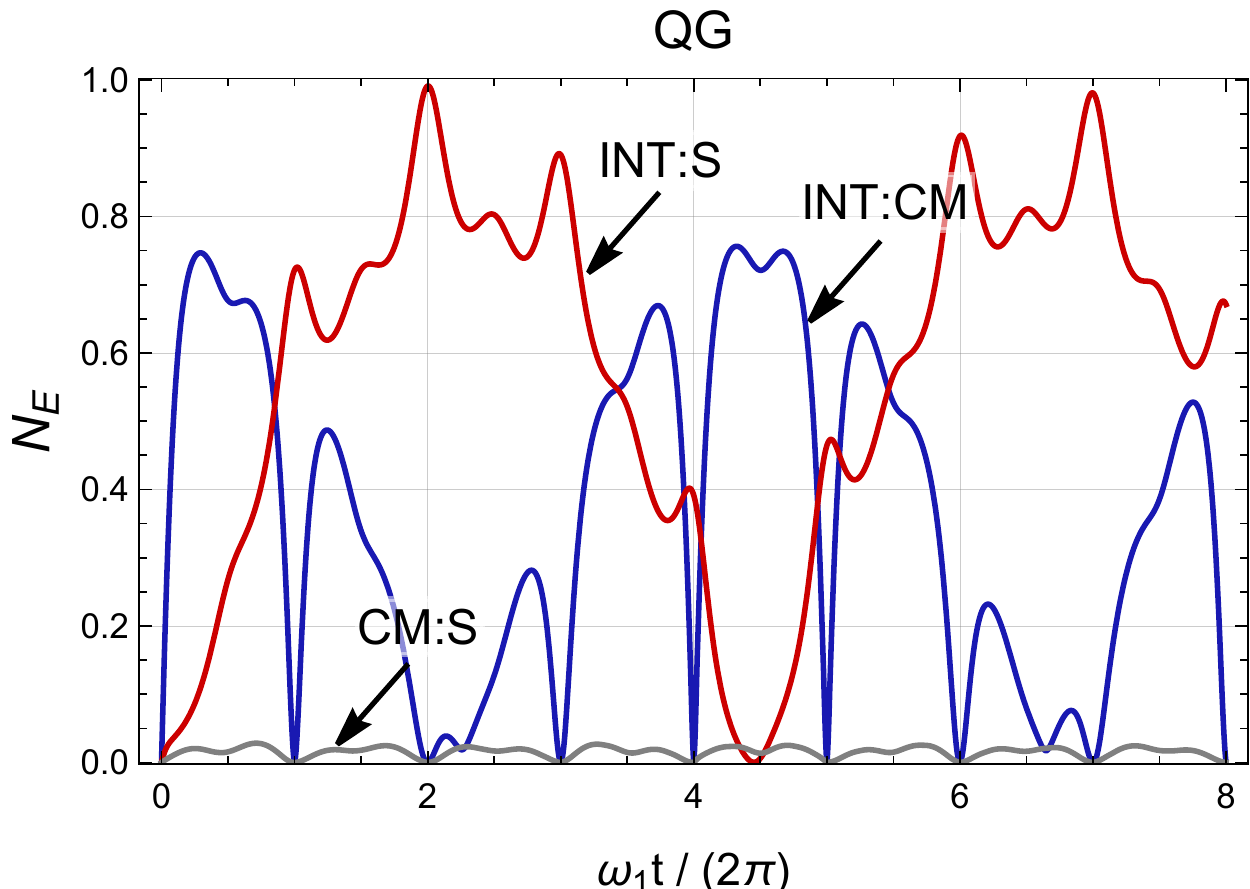}
    \caption{Logarithmic negativity of reduced bipartite states for CG case (left panel) and QG
      case (right panel). Bipartite subsystems are denoted as CM:INT
      (blue line), INT:S (red line) and CM:S (gray
      line).}
      \label{fig:lognega}
\end{figure}

Figure~\ref{fig:lognega} displays the time dependence of the
logarithmic negativity $N_E(t)$ of the reduced bipartite states for
the CG and QG cases.  The blue, red, and gray line denote entanglement of the CM:INT system, the INT:S system, and CM:S systems respectively. Each entanglement is obtained by evaluating
Eqs.~\eqref{eq:state_INTCM},\eqref{eq:state_INTS}, and
\eqref{eq:state_CMS}. The parameters are identical to
those used in Fig.~\ref{fig:visibility_CGQG}, namely
$(m_0^3/k)^{1/2}=10,~(m_0/m_1)^{1/2}=0.5,~Gm_0M/(kd^3)=0.015,~m_0^{1/4}d=10,~\beta/d=0.01,~\sigma/d=0.001$.
For the CG case, the CM:INT entanglement emerges and disappears for
every $2\pi/\omega_1$ period, and the CM:S and S:INT entanglements remain
zero. Comparing the CM:INT entanglement with the visibility in
Fig.~\ref{fig:visibility_CGQG} reveals that they both oscillate alternatively
with the $2\pi/\omega_1$ period. Therefore, the CM:INT entanglement induced by the
relativistic and classical gravitational redshift is reflected in the
revival of visibility in the Ramsey interference.  For the QG case,
whole
the CM:INT, INT:S, and CM:S entanglements emerge as
time evolves. The CM:S entanglement is relatively
smaller than others, because the CM:S entanglement is obtained from the
second order of Taylor expansion in Eq.~\eqref{eq:potential_expansion}, whereas the others originated from the first order. An important effect of the quantumness of gravity appears in the INT:S entanglement which dominates the CM:INT entanglement alternatively
during the time evolution. As the source state
$\ket{\varphi_{\pm\beta}}$ reduces to a two-level state for
$\sigma\to0$, the INT:S entanglement can achieve nearly the maximal
logarithmic negativity value of one as the reduced state nearly evolves to the
Bell state. Moreover, the beat of the visibility in
Fig.~\ref{fig:visibility_CGQG}  corresponds exactly to the envelope of
the oscillation of the INT:S entanglement. If the  CM:S
entanglement is neglected, we can conclude that the nonrevival property of the
Ramsey interference exactly reflects the creation of the INT:S entanglement revealing the quantumness of gravity.

\begin{figure}[H]
    \centering
    \includegraphics[width=1\linewidth,clip]{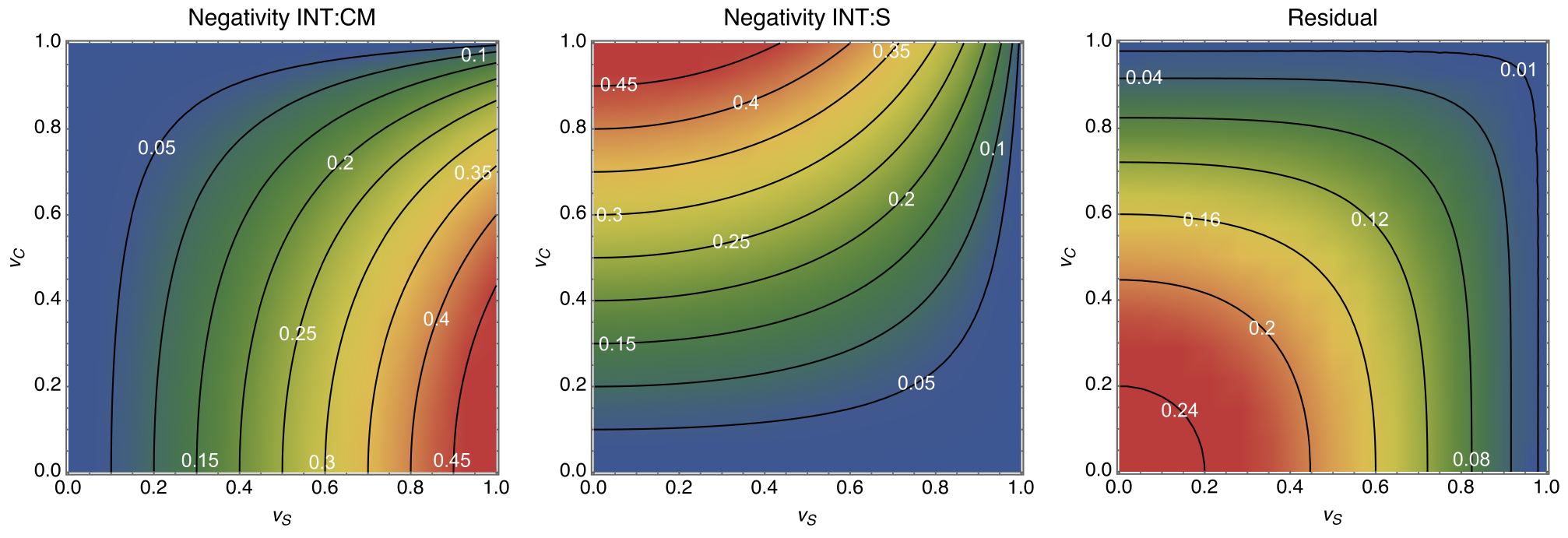}
    \caption{Dependence of negativity on $v_S$ and $v_C$ for reduced
      bipartite states. Left panel: INT:CM. Middle panel: INT:S. Right
      panel: residual entanglement, which quantifies genuine
      tripartite entanglement.  For the CG case, since $v_S=1$,
      $\mathcal{N}(\rho_\text{INT:S})$ is always zero, but
      $\mathcal{N}(\rho_\text{INT:CM})\neq 0$  depending
       on the value of $v_C$. }
    \label{fig:lognega1}
\end{figure}

As per Fig.~\ref{fig:lognega}, CM:S entanglement is negligible under
the condition that the Taylor expansion in
Eq.~\eqref{eq:potential_expansion} is validated. Now, 
we again consider up to the first order of the Taylor expansion in
Eq.~\eqref{eq:potential_expansion}  to neglect the
CM:S entanglement and derive an explicit analytic form of negativity
for three different reduced bipartite states to achieve further qualitative understanding.  In this approximation, since the dependence of $X$
only appears in $\mathcal{E}_j$ [See also \eqref{eq:AB_firstorder}],
Eq.~\eqref{eq:Psi} becomes
\begin{align}
    \ket{\Psi(t)}&=
    \frac{1}{\sqrt{2}}\sum_{j=0,1}\kt{E_j}\otimes\kt{\tilde
                   C_j}\otimes
                   \kt{\tilde S_j},
                   \label{eq:state3}
\end{align}
where
\begin{align}
    \kt{\tilde C_j}:=\kt{\psi_j(t)}
    ,\quad
    \kt{\tilde S_j}:=\int dX\, e^{-i\mathcal{E}_j(X)t}\,\varphi_S(X)\kt{X}.
\end{align}
In particular, in the qubit limit of the CM state and the source state
with
$\br{\tilde C_j}\tilde C_k\rangle\rightarrow\delta_{jk}, \br{\tilde S_j}\tilde
S_k\rangle\rightarrow\delta_{jk}$, we obtain the Greenberger-Horne-Zeilinger
(GHZ) state for the total system as
$\kt{\Psi_{\text{GHZ}}}=\left(\kt{000}+\kt{111}\right)/\sqrt{2}$.
Therefore, the existence of genuine
tripartite entanglement for the CM-INT-S state is confirmed.  To
evaluate the negativity for the reduced states,  orthonormal basis
$\kt{c_j}$ for the CM state and $\kt{s_j}$ for the S state, which
satisfy $\bra{c_j}\ket{c_k}=\bra{s_j}\ket{s_k}=\delta_{jk}$ are
introduced as follows:
\begin{align}
   &\begin{bmatrix}
      \kt{\tilde C_0}\\
      \kt{\tilde C_1}
    \end{bmatrix}
    =\frac{1}{\sqrt{2}}
    \begin{bmatrix}
      e^{i\theta_C/2}\sqrt{1+v_C} & e^{i\theta_C/2}\sqrt{1-v_C}\\
      e^{-i\theta_C/2}\sqrt{1+v_C} & -e^{-i\theta_C/2}\sqrt{1-v_C}
    \end{bmatrix}
    \begin{bmatrix}
      \kt{c_0}\\
      \kt{c_1}
    \end{bmatrix}
    ,\quad
    v_C=\left|\br{\tilde C_0}\tilde C_1\rangle\right|
    ,\quad
    \theta_C=\arg\left[\br{\tilde C_0}\tilde C_1\rangle\right], \\
   & \notag \\
   &\begin{bmatrix}
      \kt{\tilde S_{0}}\\
      \kt{\tilde S_{1}}
    \end{bmatrix}
    =\frac{1}{\sqrt{2}}
    \begin{bmatrix}
      e^{i\theta_S/2}\sqrt{1+v_S} & e^{i\theta_S/2}\sqrt{1-v_S}\\
      e^{-i\theta_S/2}\sqrt{1+v_S} & -e^{-i\theta_S/2}\sqrt{1-v_S}
    \end{bmatrix}
    \begin{bmatrix}
      \kt{s_0}\\
      \kt{s_1}
    \end{bmatrix}
    ,\quad
    v_S=\left|\br{\tilde S_0}\tilde S_1\rangle\right|
    ,\quad
    \theta_S=\arg\left[\br{\tilde S_0}\tilde S_1\rangle\right].
\end{align}
Using the orthonormal basis $\{\ket{E_0},\ket{E_1},\ket{c_0},
  \ket{c_1}, \ket{s_0}, \ket{s_1}\}$, the total
  state is expressed as
  \begin{equation}
    \rho=\ket{\Psi}\bra{\Psi}=\frac{1}{8}
    \begin{bmatrix}
      \mathbf{N}_C\otimes\mathbf{N}_S &
      e^{i(\theta_C+\theta_S)}\,\mathbf{N}_C\,\mathbf{Z}\otimes\mathbf{N}_S\,\mathbf{Z}\\
      e^{-i(\theta_C+\theta_S)}\,\mathbf{Z}\,\mathbf{N}_C\otimes\mathbf{Z}\,\mathbf{N}_S&
      \mathbf{Z}\,\mathbf{N}_C\,\mathbf{Z}\otimes\mathbf{Z}\,\mathbf{N}_S\,\mathbf{Z}
    \end{bmatrix},
    \label{eq:rho}
  \end{equation}
  where the $2\times 2$ sub-matrices are defined by
  \begin{equation}
    \mathbf{N}_C=
    \begin{bmatrix}
      1+v_C & \sqrt{1-v_C^2}\\
      \sqrt{1-v_C^2} & 1-v_C
    \end{bmatrix}
    ,\quad
     \mathbf{N}_S=
    \begin{bmatrix}
      1+v_S & \sqrt{1-v_S^2}\\
      \sqrt{1-v_S^2} & 1-v_S
    \end{bmatrix}
    ,\quad
    \mathbf{Z}=
    \begin{bmatrix}
      1 & 0\\
      0 & -1
    \end{bmatrix}.
  \end{equation}
  Therefore, the three different reduced bipartite states are
  \begin{align}
        \rho_{\text{INT:CM}}
    =\mathrm{Tr}_\text{S}\rho&=\frac{1}{4}
    \begin{bmatrix}
      \mathbf{N}_C & v_S\, e^{i(\theta_C+\theta_S)}\,\mathbf{N}_C\,\mathbf{Z}\\
      v_S\,e^{-i(\theta_C+\theta_S)}\,\mathbf{Z}\,\mathbf{N}_C &
      \mathbf{Z}\,
      \mathbf{N}_C\,\mathbf{Z}
    \end{bmatrix},
   \label{eq:matrix_INCM}  \\
    \rho_{\text{INT:S}}
    =\mathrm{Tr}_\text{CM}\rho&=\frac{1}{4}
    \begin{bmatrix}
      \mathbf{N}_S & v_c\,e^{i(\theta_C+\theta_S)}\,\mathbf{N}_S\mathbf{Z}\\
      v_c\,e^{-i(\theta_C+\theta_S)}\,\mathbf{Z}\,\mathbf{N}_S &
      \mathbf{Z}\,\mathbf{N}_S\,\mathbf{Z}
    \end{bmatrix},
      \label{eq:matrix_INS}           \\
    \rho_\text{CM:S}=\mathrm{Tr}_\text{INT}\rho&=\frac{1}{8}\left(\mathbf{N}_C\otimes\mathbf{N}_S+\mathbf{Z}\,\mathbf{N}_C\,\mathbf{Z}\otimes\mathbf{Z}\,\mathbf{N}_S\,\mathbf{Z}\right). \label{eq:matrix_CMS}
  \end{align}
 As per Eq. \eqref{eq:matrix_CMS}, $\rho_\text{CM:S}$ is separable from its
  structure. The negativity for reduced bipartite states are
\begin{align}
  \mathcal{N}(\rho_{\text{INT:CM}})&=\frac{1}{4}
  \left(-1+v_S+\sqrt{(1+v_S)^2-4v_C^2\,v_S}\right), \label{eq:N_INTCM}\\
   \mathcal{N}(\rho_{\text{INT:S}})&=\frac{1}{4}
    \left(-1+v_C+\sqrt{(1+v_C)^2-4v_S^2\,v_C}\right),
                                     \label{eq:N_INTS} \\
  \mathcal{N}(\rho_\text{CM:S})&=0.
\end{align}
Figure~\ref{fig:lognega1} displays the dependence of negativities
$\mathcal{N}(\rho_\text{INT:CM})$, $\mathcal{N}(\rho_\text{INT:S})$ on
$v_S$ and $v_C$. For the CG case, since the state is
not dependent on $X$, we have $v_S=1$. Therefore,
$\mathcal{N}(\rho_\text{INT:S})$ becomes zero, whereas
$\mathcal{N}(\rho_\text{INT:CM})$ varies between 0 and 1 depending on
values of $v_C$.  For the QG case, since $v_S$ oscillates between 0 and
1, $\mathcal{N}(\rho_\text{INT:CM})$ oscillates between $0$ and
$\sqrt{1-v_C^2}/2$, whereas $\mathcal{N}(\rho_\text{INT:S})$ oscillates
between $0$ and $v_C/2$; $\mathcal{N}(\rho_\text{INT:CM})$ and
$\mathcal{N}(\rho_\text{INT:S})$ vary alternatively. This implies a 
monogamous relation of quantum entanglement between the INT:CM systems
and INT:S systems, and confirms the  existence of genuine tripartite
entanglement of the CM-INT-S system.

For the state expressed in Eq. \eqref{eq:rho}, the negativity for the
bipartition INT:CM+S of the total state is
  \begin{equation}
   \mathcal{N}(\rho_\text{INT:CM+S})=\frac{1}{2}\sqrt{1-v_S^2\,v_C^2}, 
  \end{equation}
  and it obeys  monogamy inequality \cite{Ou2007} as follows:
  \begin{equation}
    \mathcal{N}^2(\rho_\text{INT:CM})+
    \mathcal{N}^2(\rho_\text{INT:S})\le  \mathcal{N}^2(\rho_\text{INT:CM+S}).
  \end{equation}
  The difference between the sides of this inequality represents the
  residual entanglement which quantifies genuine tripartite
  entanglement (right panel of Fig. \ref{fig:lognega1}). For the CG
  case, since $v_S=1$, there is no residual entanglement. For the QG
  case, since $v_C\neq 1$ or $v_S\neq 1$, the value of residual
  entanglement is nonzero. The point $v_C=v_S=0$ corresponds to the
  maximally entangled GHZ state.

\begin{figure}[H]
    \centering
    \includegraphics[width=0.45\linewidth,clip]{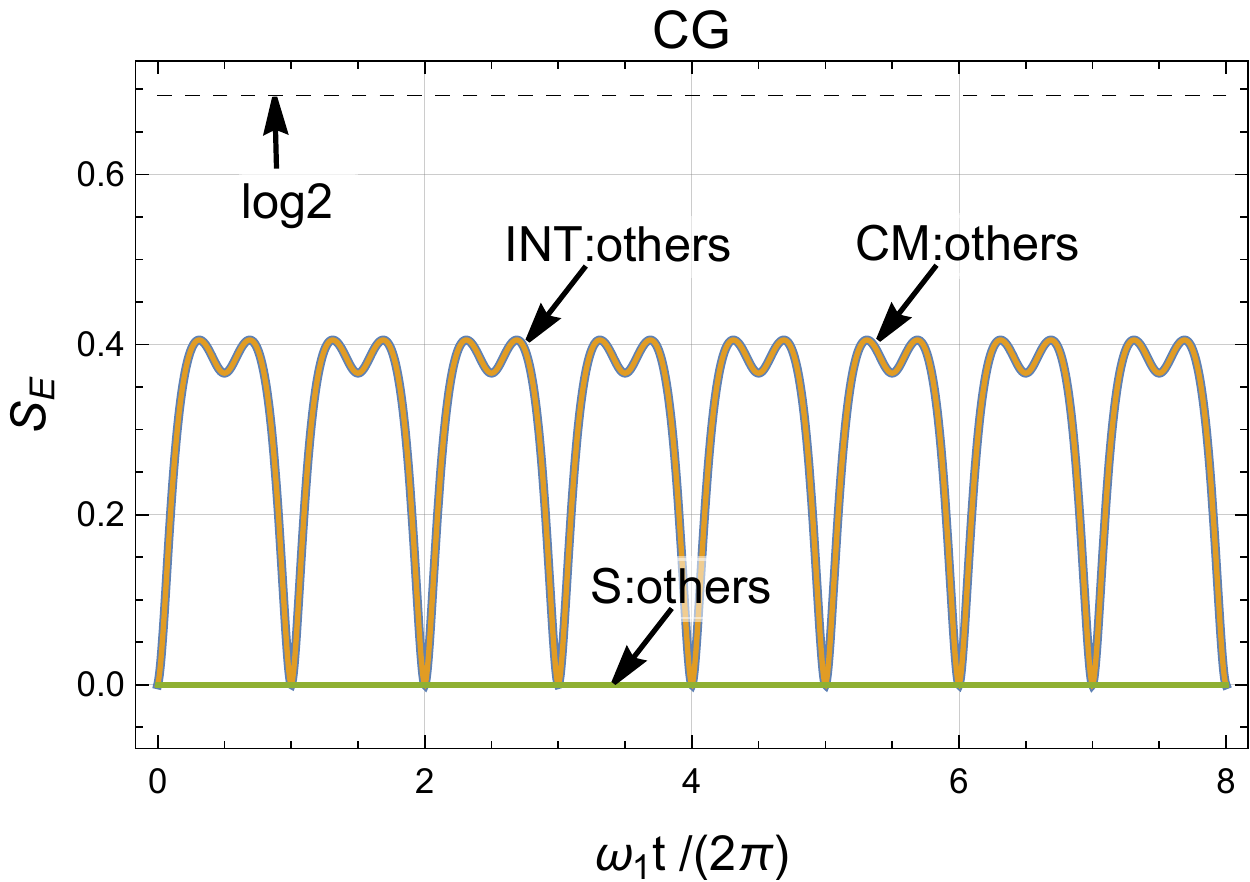}
    \hspace{0.5cm}
    \includegraphics[width=0.45\linewidth,clip]{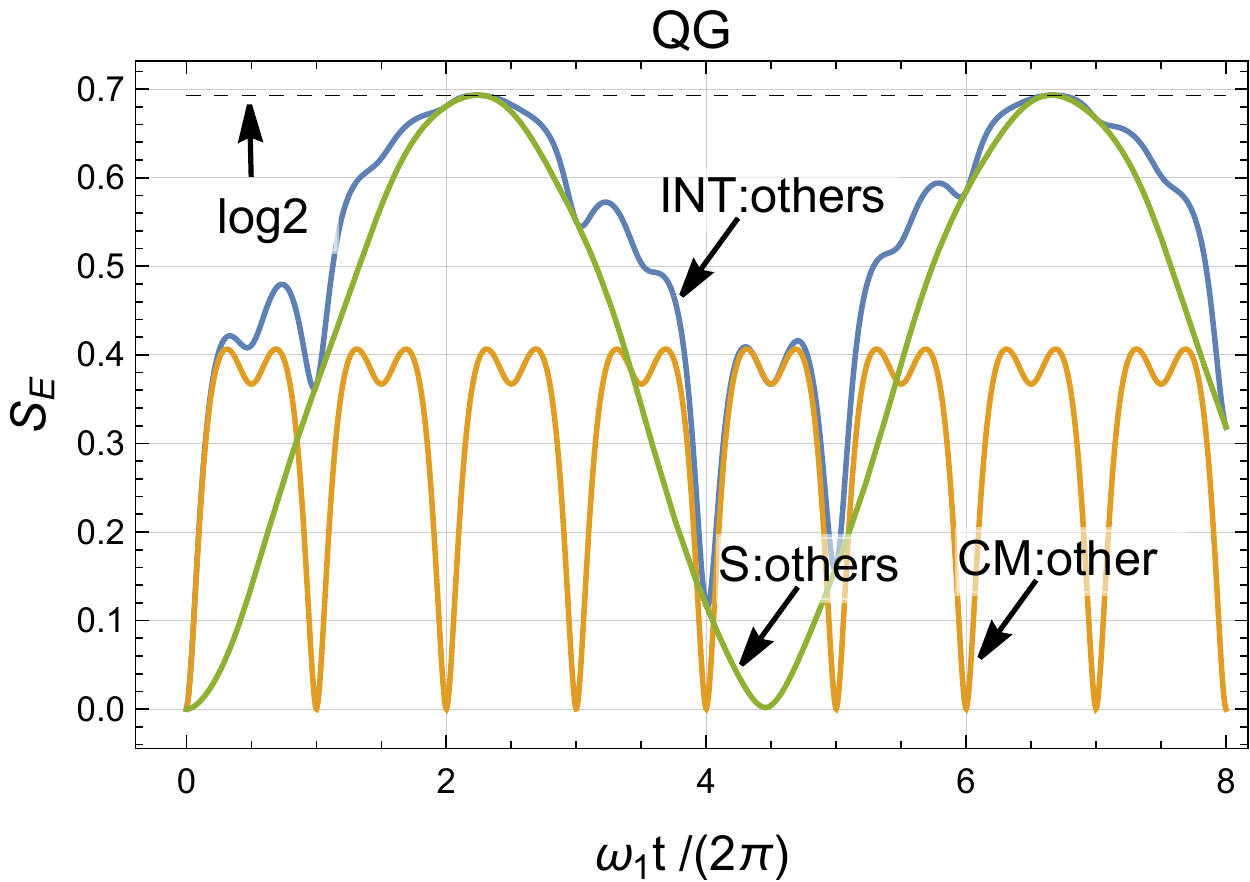}
    \caption{
    Entanglement entropy between unipartite-bipartite systems for CG case (left panel) and QG case (right panel). Each line shows CM:others systems (orange line), INT:others systems (blue line), and S:others systems (green line). The gray dashed line shows $\log 2$; the maximally entangled state.}
    \label{fig:EE}
\end{figure}

Finally, we evaluated the entanglement entropy in our system. In Fig.~\ref{fig:EE}, we depicted the time dependence of the entanglement entropy between unipartite-bipartite systems. The left panel shows the CG case, and the right panel shows the QG case. The orange line shows CM:others systems, the blue line shows INT:others systems and the green line shows the S:others systems. Parameters are chosen as the same as in Fig.~\ref{fig:visibility_CGQG}, namely $(m_0^3/k)^{1/2}=10,~(m_0/m_1)^{1/2}=0.5,~Gm_0M/(kd^3)=0.015,~m_0^{1/4}d=10,~\beta/d=0.01,~\sigma/d=0.001$. 
We see any entanglement entropy is bounded by $\log 2$ (gray dashed line), which value involves the GHZ state. In the CG case, We can see that the entanglement entropy of S:others systems vanishes. This means that the source system is isolated from the particle system due to the classical interaction. The entanglement entropy of CM:others and INT:others systems take the same value since they both show the entanglement shared between the CM system and the INT system. 
In the QG case, we can see the entanglement is shared between the whole 3 systems. Especially, the entanglement entropy of INT:others and S:others almost reach the maximum value $\log 2$ when the visibility is at the minimum in Fig.~\ref{fig:visibility_CGQG}. This is because the INT system is represented by 2 qubits and the S state is approximately represented by 2 qubits which nearly reads to the maximally entangled state.

\section{Discussion}
\label{sec:Discussion}
In this section, we discuss the feasibility of detecting the
quantumness of gravity in our proposal. Let us suppose an experiment
being performed using an $^{27}\mathrm{Al}^+$ quantum clock
\cite{Brewer2019} with a probe laser wavelength
$\lambda=2\pi \hbar c/E=267\,\mathrm{nm}$. A coherent state of the
mesoscopic mass source with $M\sim 1\,\mathrm{ng}$ may be realized in
the near future. In addition, the coherent state can be experimentally
realized for $20\,\mathrm{sec}$ \cite{Xu2019}. Apart from that, we
assume $d=200\,\mu\mathrm{m},~\sigma=1\,\mu\mathrm{m}$ and
$\beta=10\,\mu\mathrm{m}$ to compare setups of the BMV proposal
\cite{Bose2017,Marletto2017,Christodoulou2019}. Therefore, for a
duration of coherence time scale $t$, the fractional change of the
decoherence factor in $|\mathcal{V}_Q(t)|$ of Eq.~\eqref{eq:VQ} can be
estimated as
\begin{align}
    &\left(\frac{1}{2\hbar}\frac{GME}{d}\frac{\sigma}{d}\,t\right)^2
  =1.7\times 10^{-34} \left(\frac{M}{10\,\mathrm{ng}}\right)^2
  \left(\frac{\lambda}{267\,\mathrm{nm}}\right)^{-2}
  \left(\frac{d}{200\,\mu\mathrm{m}}\right)^{-4}
  \left(\frac{\sigma}{1\,\mu\mathrm{m}}\right)^2\left(\frac{t}{20\,\mathrm{sec}}\right)^2,\\
    &\left(\frac{1}{\hbar}\frac{GME}{d}\frac{\beta}{d}\,t\right)^2
  =6.8\times 10^{-32} \left(\frac{M}{10\,\mathrm{ng}}\right)^2
  \left(\frac{\lambda}{267\,\mathrm{nm}}\right)^{-2}
  \left(\frac{d}{200\,\mu\mathrm{m}}\right)^{-4}
  \left(\frac{\beta}{10\,\mu\mathrm{m}}\right)^2\left(\frac{t}{20\,\mathrm{sec}}\right)^2.
  \label{eq:our-result}
\end{align}
After performing time Fourier transformation on the probability $P(t)$
obtained using the spectroscopy experiments, the least necessary
precision to detect the QG effect in our proposal is approximately
$10^{-32}$, which is extremely small to be distinguished in
the present clock spectroscopy, whose observation uncertainty is
about $10^{-19}$ \cite{Brewer2019}. Let us discuss our result Eq. \eqref{eq:our-result} in comparison with visibility change
obtained by the setup of Carney \textit{et al.} \cite{Carney2021}
. Carney \textit{et al.} investigated entanglement
between a massive oscillator and a source mass particle with a cat
state, and evaluated the interference visibility of the particle
state. Their estimation of the time change of the visibility due to
quantum gravitational interaction is
\begin{equation}
  \left(\frac{GMm}{d}\frac{\beta}{d}\frac{x_0}{d}\,t\right)^2\propto d^{-6},
  \label{eq:carney}
\end{equation}
where $x_0$ is the spread of the ground state wave function of the oscillator.
For the optimal values of parameters in their setup, the
ratio provides a value of $\sim 10^{-28}$. Although
Eq. \eqref{eq:our-result} exhibits a lower suppression factor $d^{2}$
compared to \eqref{eq:carney}, the ratio  $E/m$ makes the visibility
change smaller compared to that in Eq. \eqref{eq:carney}.

\begin{figure}[H]
  \centering
  \includegraphics[width=0.7\linewidth,clip]{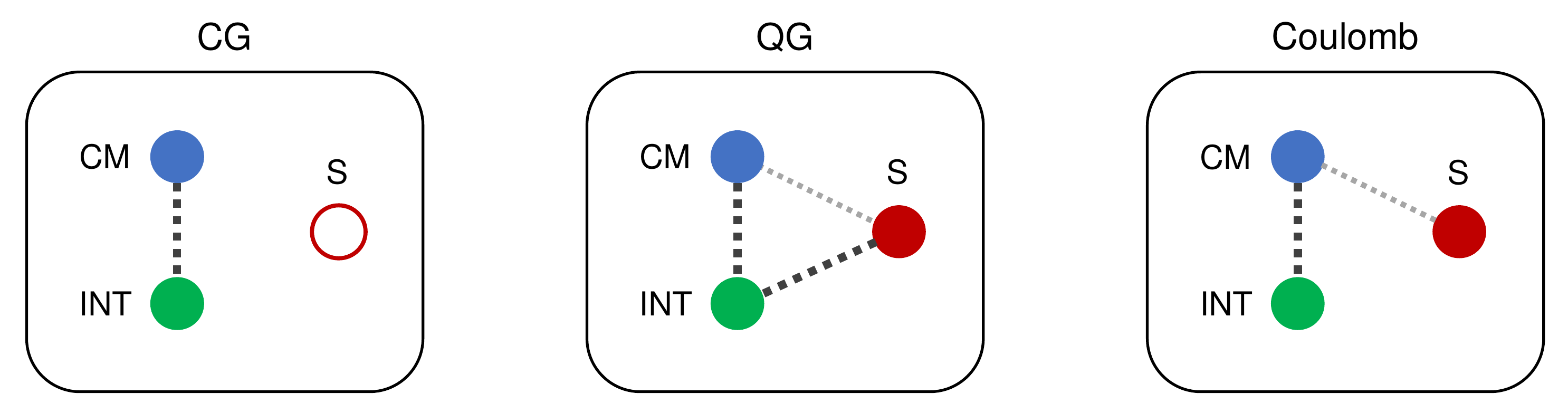}
  \caption{Entanglement structure for the  CG, QG, and Coulomb interaction
    cases. Dotted lines represent the existence of couplings between
      degrees of freedom, and possible pairs of subsystems sharing
      entanglement. For the QG case, the total system shares genuine
      tripartite entanglement which cannot be reduced to bipartite
      entanglement between the reduced bipartite systems. For the quantum
      Coulomb case, tripartite entanglement is reduced to
      bipartite entanglement between the reduced bipartite systems.}
    \label{fig:structure}
\end{figure}

\begin{figure}[H]
    \centering
    \includegraphics[width=0.45\linewidth,clip]{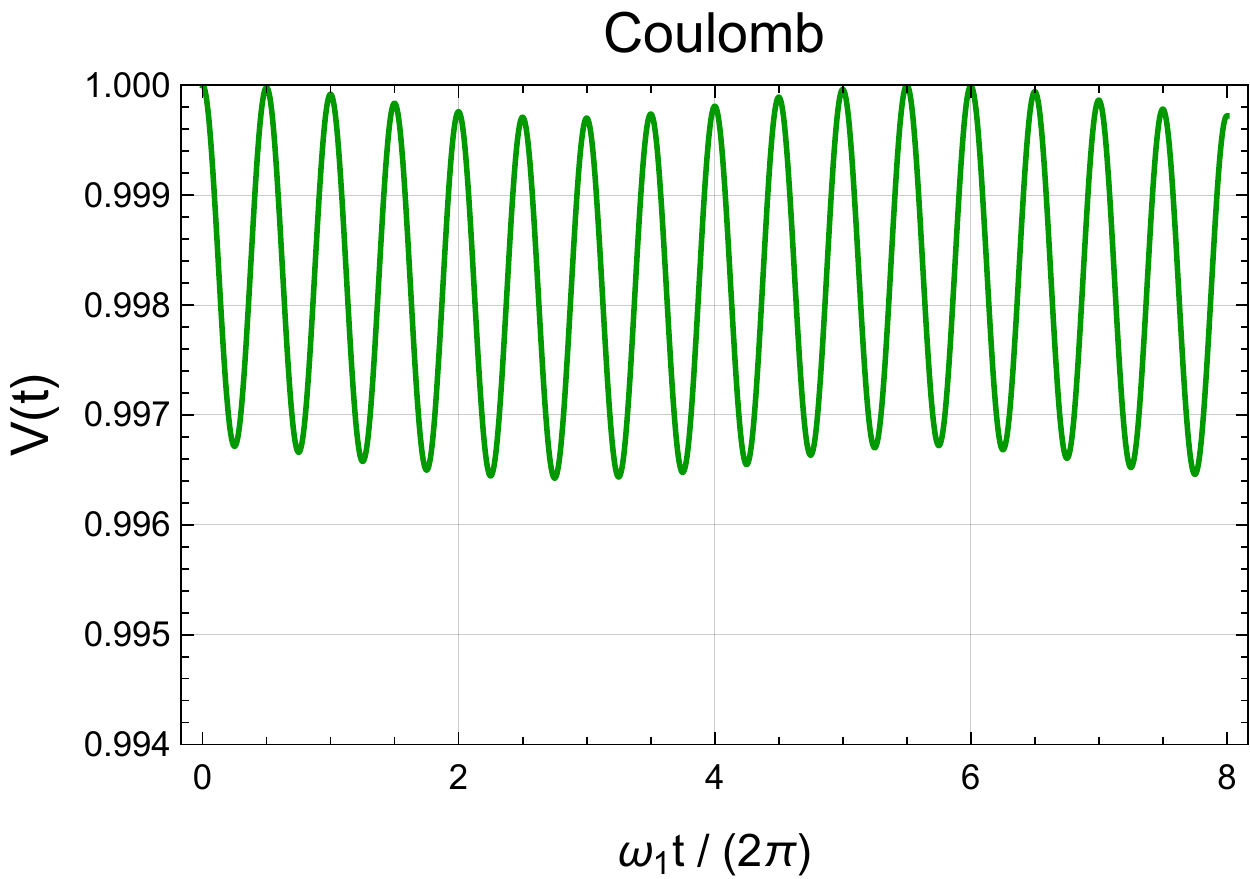}
   \hspace{0.5cm}
    \includegraphics[width=0.45\linewidth,clip]{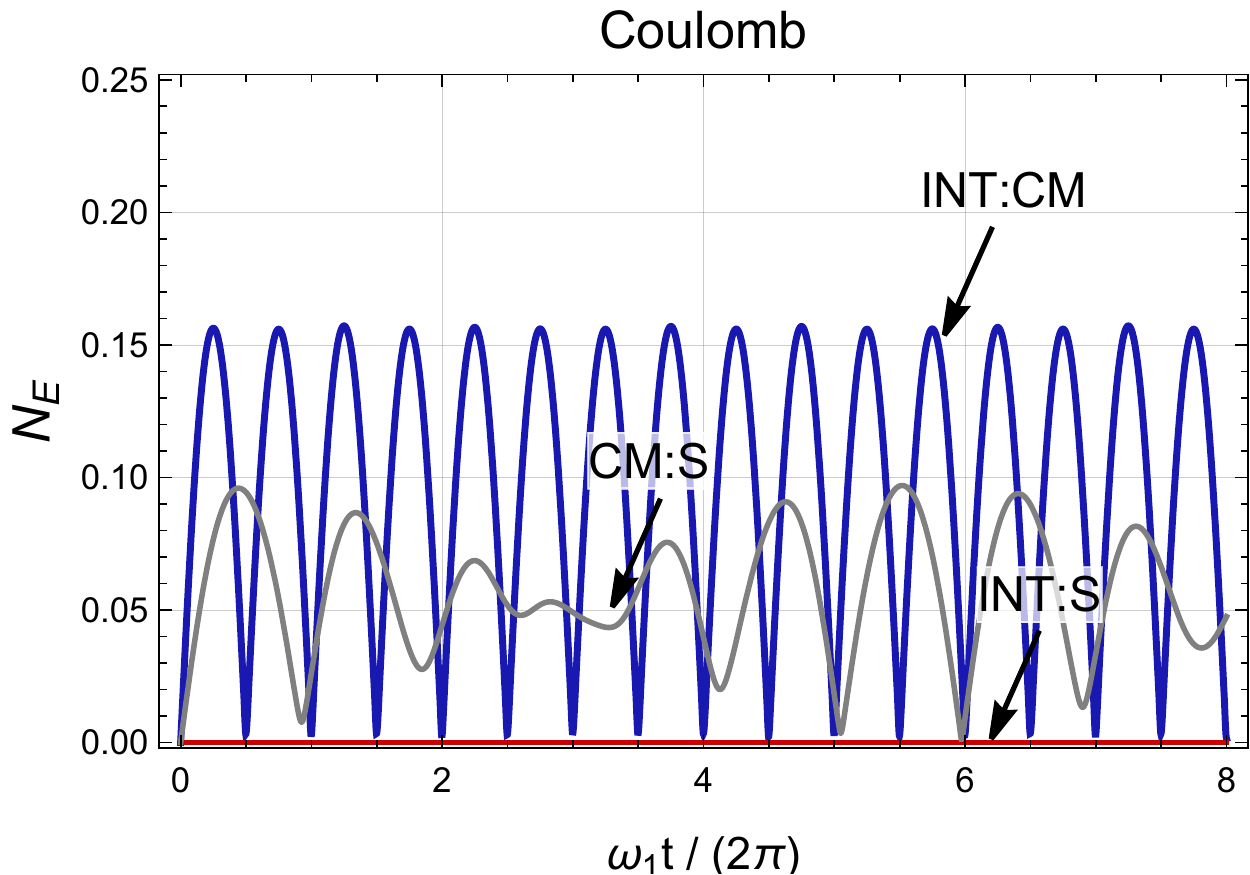}
    \caption{Left panel: time dependence of the visibility $\mathcal{V}(t)$ for
      the Coulomb case.  Right panel: logarithmic negativity for the
      Coulomb case; Red line: bipartite INT:S subsystems. Blue line:
      bipartite INT:CM subsystem. Grayline: bipartite CM:S subsystem.}
    \label{fig:probability_Coulomb}
\end{figure}

As a crucial issue of this study, we also discuss the uniqueness of
gravity compared to other quantum interactions, such as the
electromagnetic Coulomb force.  In our setup, CG creates an
entanglement only between the CM and INT systems, whereas all
subsystems share entanglement in the QG case. It should be noted
that, in particular, the entanglement between the INT and S systems is
uniquely induced by the quantumness of gravity, and no other quantum
interactions can establish this entanglement. This is because according to the weak equivalence
principle and the mass-energy equivalence, only gravity can couple to
the energy; other quantum interactions do not have this property
\cite{Pikovski2015}. Figure~\ref{fig:structure} depicts the
entanglement structure for different external forces, namely CG, QG,
and Coulomb force. To capture the uniqueness of the quantumness of
gravity, we should detect entanglement held between the INT system and
the S system.
For comparison, let us consider the case when the Coulomb interaction
is the external force instead of the gravitational interaction. To
focus on the connection of mass-energy equivalence with entanglement here,
we do not consider a dipole-photon interaction which introduces a
direct coupling between the INT and S system.\footnote{
  Note that we have made an assumption in our setup that the internal degrees of freedom of the particle are labeled only by the energy levels, which results in the unique correspondence of the INT:S entanglement and QG. Generally, the internal degrees of freedom of an atom are labeled not only by energy level but also by spin. In such cases, the dipole-photon interaction can also create INT:S
  entanglement besides gravity. A general form of the dipole-photon
  interaction is given by $\hat{\mathbf{d}}\cdot\hat{\mathbf{E}}$,
  where $\hat{\mathbf{d}}=d_+\ket{E_1}\bra{E_0}+d_-\ket{E_0}\bra{E_1}$ is the
  electric dipole moment operator of the INT system, and
  $\hat{\mathbf{E}}$ is the external electric field. When the external
  electric field has the source (S) dependence as
  $\hat{\mathbf{E}}(\hat X)$, this interaction can provide the INT-S
  entanglement in the leading order of expansion for the
  source separation.} 
The Coulomb interaction is given by
\begin{align}
    V_{\text{Coulomb}}=\frac{q Q}{4\pi\epsilon_0 |d+\hat X-\hat x|},
\end{align}
where $q$ and $Q$ are the electric charges of the oscillator and the
source, respectively. In comparison with the QG interaction
\begin{align}
    V_{\text{QG}}=-\frac{G\hat m M}{|d+\hat X-\hat x|},
\end{align}
the Coulomb interaction does not contain the INT operator $\hat m$ as a result of the weak equivalence principle. Then, it is obvious that the Coulomb interaction does not produce the entanglement between the INT and S systems as depicted in Fig.~\ref{fig:structure}.
The Hamiltonian for the external Coulomb interaction can be
expressed as 
\begin{align}
  &\hat H_\text{Coulomb}(\hat X)=
    \sum_{j=0,1}\left[\mathcal{E}_j(\hat X)
    +\frac{\hat p^2}{2m_j}+\frac{m_j\omega_j^2}{2}
    \left(\hat x-\frac{qB(\hat X)}{k}\right)^2\right]\ket{E_j}\bra{E_j},\quad
  \mathcal{E}_j(\hat X)=m_j-qA(\hat X)-\frac{q^2B^2(\hat
    X)}{2k}, \\
  &A(\hat X)=\frac{Q}{4\pi\epsilon_0 d}\left(1-\frac{\hat X}{d}+\frac{\hat
    X^2}{d^2}\right),\quad
    B(\hat X)=\frac{Q}{4\pi\epsilon_0 d^2}\left(1-\frac{2\hat X}{d}\right),\quad
    C=\frac{Q}{4\pi\epsilon_0 d^3}.
\end{align}
Unlike the gravity case, the symmetry axis of the trapping potential is independent of the INT
state, and the quantity
$\mathcal{E}_1(\hat X)-\mathcal{E}_0(\hat X)$ is independent of $\hat
X$. 
As a result, the visibility of the Coulomb force case exhibits a 
different behavior compared to the CG and QG cases.
Following the same procedure used to derive Eq.~\eqref{eq:V_QG},
we obtain the $\pi/\omega_1$ periodic visibility as
\begin{align}
    |\bra{\Psi_0(t)}\ket{\Psi_1(t)}|=\left(\frac{4a_0\,a_1}{4a_0\,a_1\cos^2(\omega_1t)+(a_0+a_1)^2\sin^2(\omega_1 t)}\right)^{1/4}.
    \label{eq:V1_Coulomb}
\end{align}
Here, the period $2\pi/\omega_1$ of coherent state does not appear
since the displacement factor
$\Delta(\hat X):=qB(\hat X)/k\approx Q/d^2$ is independent of the internal energy level.  Therefore, up to the leading order in Eq.~\eqref{eq:potential_expansion}, we conclude that the
nonrevival behavior in visibility uniquely discriminates the
quantumness of gravity.  The visibility is more likely to exhibits
revival behavior for the Coulomb interaction case than the QG case
since the former shares less entanglement than the latter, as in
Fig.~\ref{fig:structure}.  The left panel of
Fig.~\ref{fig:probability_Coulomb} presents the behavior of the
probability for the Coulomb interaction by evaluating up to the
second-order of Taylor expansion in
Eq.~\eqref{eq:potential_expansion}. The parameters are set to
$(m_0^3/k)^{1/2}=10,~(m_0/m_1)^{1/2}=0.85,~qQ/(4\pi\epsilon_0
kd^3)=0.15,~m_0^{1/4}d=10,~\beta/d=0.01,~\sigma/d=0.001$. As the figure shows, the visibility exhibits a revival behavior with some longer period
than $\pi/\omega_1$ obtained by the first-order approximation
in Eq.~\eqref{eq:V1_Coulomb}. This is because the visibility beats due
to the addition of two functions with periods $2\pi/\omega_0$ and
$2\pi/\omega_1$,  reflecting the respective period of the coherent states
$\ket{\Psi_0(t)}$ and $\ket{\Psi_1(t)}$ as an effect of the external
quantum Coulomb force.  In the first-order
approximation, the displacement effects of these coherent states are
neglected. Regarding the  behavior of negativity, within the first order
approximation, $v_S=1$ results in $\mathcal{N}(\rho_\text{INT:S})=0$
and provides a nonzero value of $\mathcal{N}(\rho_\text{CM:INT})$
with period $\pi/\omega_1$. The quantumness of Coulomb force appears as
a small nonzero value of $\mathcal{N}_\text{CM:S}$, which can be
revealed beyond the first-order approximation, as per the right panel of
Fig. \ref{fig:probability_Coulomb}.

Also, we briefly discuss the case where the harmonic oscillator potential trapping the particle is proportional to the probe mass \cite{Katori2011}:
$\frac{1}{2}\hat{m}\,\omega^2\hat{x}^2$, where $\omega$ is a constant
parameter. Here, a shift of the potential due to the external gravity
becomes independent of the internal energy level, and we obtain
$\Delta_1=\Delta_0\neq 0$. Since the displacement effect does not
become obvious in the Ramsey interference, $|\mathcal{V}_C(t)|$
reduces to a $\pi/\omega_1$ periodic function rather than
$2\pi/\omega_1$ periodic function. Therefore, the visibility revival
period becomes $\pi/\omega_1$ for the CG case, while it does not
exhibit revival for the QG case due to the quantumness of gravity.

Finally, we comment on the relevance of the alternative theories of gravity. 
Since our proposal are focusing only on the nonrelativistic scale, it is difficult to distinguish alternative theories of gravity when they reduce to the Newtonian gravity in the Newtonian limit. To distinguish them, we need experiment on the higher orders of $1/c^2$ expansion where the difference between alternative theories appears. It may be interesting to discuss how the Ramsey interference differs depending on alternative theories of gravity by analyzing the higher-order expansions, although the detection is technically difficult at present.

\section{Summary}
\label{sec:Summary}

In this study, we investigated the quantumness of gravity in the setup of the Ramsey
interferometry.  For the classical external gravitational field, the
interference visibility exhibits oscillatory behavior, and the coherence
between the two energy states exhibits collapse and revival behavior in time. The decoherence behavior originated from the coupling between the CM
state and INT state due to the mass-energy equivalence principle.  On
the other hand, when a gravitational source is in quantum superposition, the visibility exhibited additional oscillation with a longer period and decay at a long time scale for the QG case, while it remains $2\pi/\omega_1$ periodic in the CG
case. For the zero spread of the source state, the time scale of
decoherence was determined as
$t\sim\left(\frac{GME}{d}\frac{\beta}{d}\right)^{-1}$, which
constrains the feasibility of the experimental detection of the quantumness of
gravity. In addition, we found that the spread of the source wave function
provides another decoherence effect that is not periodic in time depending on the variance of the source Gaussian state $\sigma$.

Regarding the entanglement behavior, the initial separable CM-INT-S
  system acquired genuine tripartite entanglement due to the quantumness
  of gravity. On the contrary, the quantumness of the Coulomb force cannot
  acquire the entanglement between the INT and S systems. This is because the Coulomb interaction
  does not couple to mass-energy of the INT state, unlike gravity,
  which obeys the weak equivalence principle. Since the nonrevival feature of the interference visibility reproduce the entanglement between the INT and S systems, it is possible to
  distinguish the quantumness of gravity from other quantum interactions by observing the
  visibility of Ramsey interference. We believe that our study is beneficial for further understanding
  of the quantum nature of gravity.

\begin{acknowledgements}
  We would like to thank A. Matsumura and K. Yamamoto for providing
  their valuable insight on the subject. This research was supported
  in part by the Japan Science and Technology Agency (JST), the Nagoya
  University Interdisciplinary Frontier Fellowship  (Y.K.), and JSPS
  KAKENHI Grant No. 19K03866 (Y.N.) and No. 22H05257 (Y.N.).
\end{acknowledgements}

\appendix
\section{WIGNER FUNCTION OF TIME EVOLVED CM STATE}
\label{appendix}

In this section we will investigate the time evolution of the CM
state $\ket{\psi_{j,X}(t)}$ by depicting its Wigner
function \cite{Wigner1932}, which gives us an intuitive understanding of
the visibility behavior.

In the QG case, the time-evolved CM state is given by
\begin{align}
    \ket{\psi_{j,X}(t)}=\hat U_{j,X}(t)\ket{\psi_{\text{ini}}},
\end{align}
where
$\hat U_{j,X}(t):=e^{-i\omega_j(\hat a_{j,X}^\dagger\,\hat
  a_{j,X}+1/2)t}$, $\ket{\psi_{\text{ini}}}$ is the ground state of
the Hamiltonian $\omega_j(\hat a_{0,0}{}^{\!\!\!\!\!\!\dagger}\,\,\,\hat a_{0,0}+1/2)$,
and $\hat a_{j,X}$ is as per Eq.~\eqref{eq:annihilation_op}.  The
relation between the two
annihilation operators $\hat a_{0,0}$ and $\hat a_{j,X}$ is as follows, by
\begin{align}
  & \hat a_{j,X}=\cosh r_j\,\hat a_{0,0}+\sinh r_j\,\hat a_{0,0}^\dagger+\alpha_{j,X},\\
  & e^{r_j}=\sqrt{\frac{m_j\omega_j}{m_0\omega_0}},\quad
  \alpha_{j,X}=\sqrt{\frac{m_0\omega_0}{2}}e^{r_j}(\Delta_0(0)-\Delta_j(X)).
\end{align}
Using the squeezing operator
$\hat S(r_j)=e^{r_j(\hat a_{0,0}^2-\hat a_{0,0}^{\dagger 2})/2}$ and
the displacement operator
$\hat D(\alpha)=e^{\alpha(\hat a_{0,0}^\dagger-\hat a_{0,0})}$, the
relation of the two annihilation operators can be rewritten as
\begin{equation}
  \hat a_{j,X}=\hat S^\dagger(r_j)\hat D^\dagger(\alpha_{j,X})\hat
  a_{0,0}\hat D(\alpha_{j,X})\hat S(r_j).
\end{equation}
Therefore,  the initial ground state $\ket{\psi_{\text{ini}}(t)}$ associated
with $\hat a_{0,0}$ evolves to become the squeezed coherent state
characterized by parameters $r_j$ and $\alpha_{j,X}$ as given below,
\begin{align}
  \ket{\psi_{j,X}(t)}=
  S^\dagger(r_j)D^\dagger(\alpha_{j,X})\hat U_{0,0}D(\alpha_{j,X})S(r_j)\ket{\psi_{\text{ini}}}.
\end{align}
For the no gravity case, the time-evolved CM state is obtained by
replacing $\alpha_{j,X}\to 0$. For the CG case, the time-evolved CM
state is obtained by replacing $\alpha_{j,X}\to \alpha_{j}$.

Next, we explore the temporal behavior of the CM Wigner
function \cite{Wigner1932}. The Wigner function is a
quasiprobability distribution in the  phase space $(x,p)$ and is defined as
\begin{align}
    W_{\psi}(x,p)=\frac{1}{2\pi}\int^\infty_{-\infty}d\xi\,
  \psi\!\left(x+\frac{\xi}{2}\right)
  \psi^*\!\left(x-\frac{\xi}{2}\right)e^{i\,\xi\, p},
\end{align}
where $\psi(x)$ is the wave function of the CM system.  For simplicity, we focus on the state
with  $X=\pm\beta$, which is the same
condition as we depicted in Fig.~\ref{fig:interf_QG}. Length unit of $2m_0\omega_0=1$.

\begin{figure}[htbp]
    \centering
    \includegraphics[width=1\linewidth,clip]{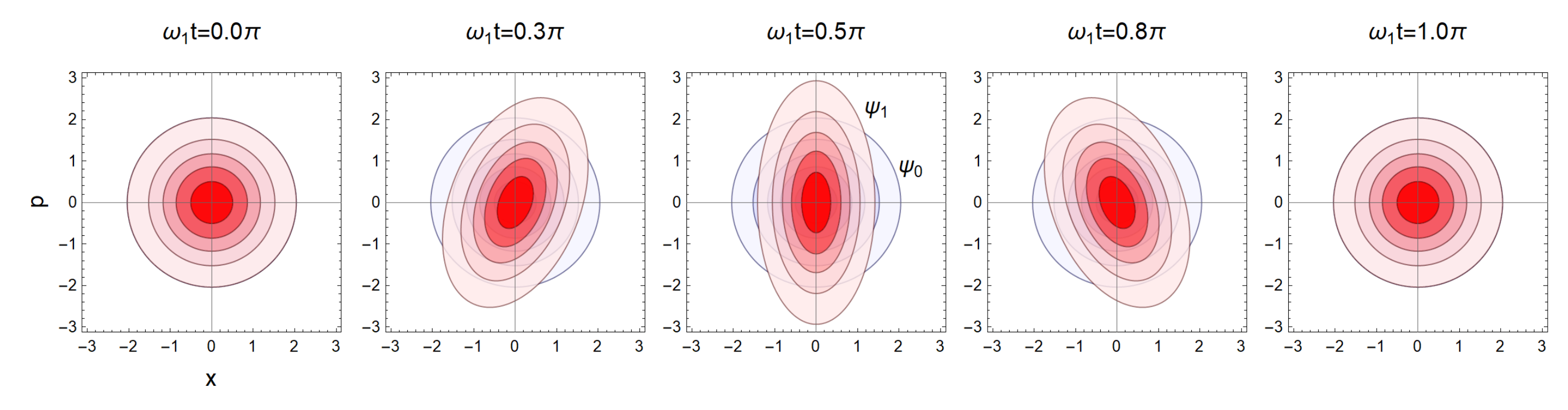}
    \caption{Time evolution of the Wigner function of the CM state for
      the no gravity case. A blue region: Wigner function of the
      $\ket{\psi_0(t)}$ state. A red region: Wigner function of the
      $\ket{\psi_1(t)}$ state. Contour lines are drawn with
      $(0.02,0.05,0.08,0.11,0.14)$.}
    \label{fig:wigner_nogravity}
\end{figure}

Figure~\ref{fig:wigner_nogravity} displays the time evolution of
the Wigner function of the CM state for the no gravity case. The blue
and red regions respectively denote the Wigner
function of the $\ket{\psi_0(t)}$ and $\ket{\psi_1(t)}$ states. The parameters are set to
$e^{r_1}=1.2,~\omega_1/\omega_0=0.8$. Since there is no displacement
effect due to gravity, the time-evolved state is simply squeezed due
to a special relativistic effect. The two states overlap for
every $\pi/\omega_1$ period, which reflects the period of squeezing. Since the
visibility of the Ramsey interference in Eq.~\eqref{eq:V1_qubit} contains
$|\bra{\psi_0(t)}\ket{\psi_1(t)}|$, its revival period $\pi/\omega_1$
stems from the squeezing period.

\begin{figure}[htbp]
    \centering
    \includegraphics[width=1\linewidth,clip]{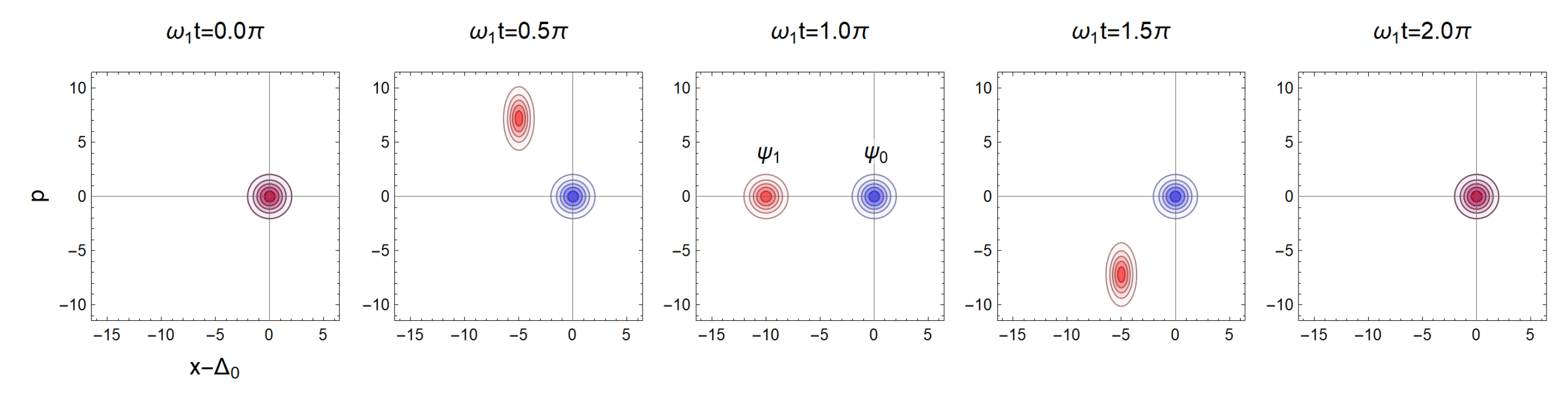}
    \caption{Time evolution of the Wigner function of the CM state for
      the CG case. A blue region: Wigner function of the
      $\ket{\psi_0(t)}$ state. A red region: Wigner function of the
      $\ket{\psi_1(t)}$ state. Contour lines are drawn with
      $(0.02,0.05,0.08,0.11,0.14)$.}
    \label{fig:wigner_CG}
\end{figure}

Figure~\ref{fig:wigner_CG} displays the time evolution of the
Wigner function of the CM state for the CG case. The blue and red
regions respectively denote the Wigner function
of $\ket{\psi_0(t)}$ and  $\ket{\psi_1(t)}$ states. The parameters are set to
$e^{r_1}=1.2,~\alpha_1=3,~\omega_1/\omega_0=0.8$. The time
evolved state is squeezed by a special relativistic effect and
displaced by a gravitational effect. The two states overlap for
every $2\pi/\omega_1$ period reflecting the period of coherent state, which
results in the $2\pi/\omega_1$ revival period of the visibility.

\begin{figure}[H]
    \centering
    \includegraphics[width=1\linewidth,clip]{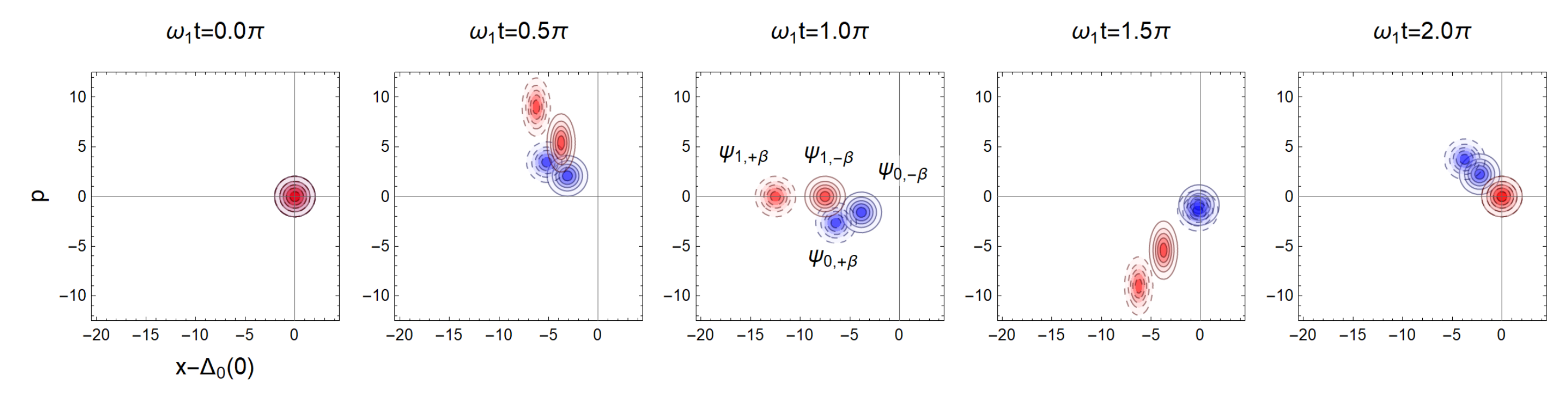}
    \caption{Time evolution of the Wigner function of the CM state
      for the QG case. Blue region with solid contours: Wigner
      function of $\ket{\psi_{0,-\beta}(t)}$. Blue region with dashed
      contours: Wigner function of $\ket{\psi_{0,+\beta}(t)}$. Red region with solid contours; Wigner function of
      $\ket{\psi_{1,-\beta}(t)}$. Red region with dashed contours:
      Wigner function of $\ket{\psi_{1,+\beta}(t)}$. Contour lines are
      drawn with $(0.02,0.05,0.08,0.11,0.14)$.}
    \label{fig:wigner_QG}
\end{figure}

Figure~\ref{fig:wigner_QG} displays the time evolution of the Wigner
function of the CM state for the CG case. The blue region with solid
contours denotes The Wigner function of $\ket{\psi_{0,-\beta}(t)}$,
the blue region with dashed contours denotes the Wigner function of
$\ket{\psi_{0,+\beta}(t)}$, the red region with solid contours denotes
the Wigner function of $\ket{\psi_{1,-\beta}(t)}$, and the red region
with dashed contours denotes the Wigner function of
$\ket{\psi_{1,+\beta}(t)}$. The parameters are set to
$e^{r_1}=1.2,~\alpha_{0,0}=1.5,~\alpha_{1,0}=3.0,~\alpha_{j,+\beta}/\alpha_{j,-\beta}=0.6,~\omega_1/\omega_0=0.8$. There
are four kinds of squeezed-coherent states with $j=0,~1$ and
$X=\pm\beta$. The Wigner functions of $\ket{\psi_{0,X}(t)}$ and
$\ket{\psi_{1,X}}$ moves with different the time periods $2\pi/\omega_0$
and $2\pi/\omega_1$, respectively, and typically do not coincide. The nonrevival behavior of the visibility for the QG case is originated
from the fact that these four states do not coincide, as well as the
phase difference $e^{-i\mathcal{E}_{j,X}t}$ in
Eq.~\eqref{eq:V1_qubit}.

%

\end{document}